\pdfoutput=1

\documentclass[11pt]{article}

\usepackage[preprint]{neurips_2023}

\usepackage[utf8]{inputenc} 
\usepackage[T1]{fontenc}    
\usepackage{url}            
\usepackage{booktabs}
\usepackage{amsfonts}       
\usepackage{nicefrac}       
\usepackage{microtype}      
\usepackage{xcolor}         

\usepackage{verbatim}       
\usepackage{wrapfig}
\usepackage{multirow}
\usepackage{gensymb}

\usepackage[export]{adjustbox} 

\usepackage{amssymb,amsfonts,amsmath,babel,mathtools,bm}

\def\vh{{\bm{h}}}

\def\vn{{\bm{n}}}

\def\vr{{\bm{r}}}

\def\vx{{\bm{x}}}

\def\mD{{\bm{D}}}

\def\mW{{\bm{W}}}

\DeclareMathAlphabet{\mathsfit}{\encodingdefault}{\sfdefault}{m}{sl}
\SetMathAlphabet{\mathsfit}{bold}{\encodingdefault}{\sfdefault}{bx}{n}

\usepackage{array}
\newcolumntype{C}{>{$\displaystyle}c<{$}}
\newcolumntype{L}{>{$\displaystyle}l<{$}}

\newcommand{\EQ}{\begin{equation}}
\newcommand{\EE}{\end{equation}}
\newcommand{\EQA}{\begin{eqnarray}}
\newcommand{\EEA}{\end{eqnarray}}

\usepackage{graphicx} 
\usepackage{amsmath}
\usepackage{algorithm2e}
\RestyleAlgo{ruled}
\SetKwComment{Comment}{/* }{ */}
\SetKwInOut{Input}{Input}\SetKwInOut{Output}{Output}

\makeatletter
    \let\@fnsymbol\@arabic
\makeatother

\usepackage{hyperref}       

\title{H-Packer: Holographic Rotationally Equivariant Convolutional Neural Network for Protein Side-Chain Packing}

\author{%
  Gian Marco Visani\thanks{Paul G. Allen School of Computer Science and Engineering, University of Washington}\\
  \And 
  William Galvin\footnotemark[1]{}
  \And
  Michael N. Pun\thanks{Department of Physics, University of Washington}
  \And
  Armita Nourmohammad\footnotemark[1]{}\,\,\,\footnotemark[2]{}\,\,\,\thanks{Department of Applied Mathematics, University of Washington}\,\,\,\thanks{Fred Hutch Cancer Research Center, Seattle, WA}
}

\begin{document}

\maketitle

\begin{abstract}

Accurately modeling protein 3D structure is essential for the design of functional proteins. An important sub-task of structure modeling is protein side-chain packing: predicting the conformation of side-chains (rotamers) given the protein's backbone structure and amino-acid sequence. Conventional approaches for this task rely on expensive sampling procedures over hand-crafted energy functions and rotamer libraries. Recently, several deep learning methods have been developed to tackle the problem in a data-driven way, albeit with vastly different formulations (from image-to-image translation to directly predicting atomic coordinates). Here, we frame the problem as a joint regression over the side-chains' true degrees of freedom: the dihedral $\chi$ angles. We carefully study possible objective functions for this task, while accounting for the underlying symmetries of the task. We propose {\em Holographic Packer} (H-Packer), a novel two-stage algorithm for side-chain packing built on top of two light-weight rotationally equivariant neural networks. We evaluate our method on CASP13 and CASP14 targets. H-Packer is computationally efficient and shows favorable performance against conventional physics-based algorithms and is competitive against alternative deep learning solutions.
    
\end{abstract}

\section{Introduction}
Proteins are macromolecules composed of residues (amino-acids) that are linked consecutively to form an amino-acid \textit{sequence}. Each residue is conceptually divided into two parts: (i) a  \textit{backbone} structure common to all amino acids, which is  comprised of the alpha carbon (C-$\alpha$) bounded to an amino group (-NH$_2$) and a carboxyl group (-COOH); and (ii) a residue-specific \textit{side-chain}. Backbones are connected by peptide bonds between the amino and carboxyl groups of consecutive residues. Physical interactions between the freely-moving side-chains cause the protein chain to \textit{fold} into a complex 3D structure, which confers the protein its function.
 
Conceptually, a protein's full atomic structure can be divided into its backbone structure (the coordinates of its backbone atoms) and its side-chains conformations (the coordinates of its side-chain atoms). Side-chain conformations are relatively flexible, while the backbone structure is more rigid and confers the protein its main 3D topology, and thus, its main function. Nonetheless, the interaction between a protein's backbone and side-chains is essential for the stability of the fold and protein function. 

Determining amino acid side-chain conformations in a protein, known as Protein Side-Chain Packing (or Rotamer Packing), is an essential step in protein folding and the de-novo design of proteins. Computational approaches to protein folding often divide the structure inference problem into two steps: first, they characterize the rigid backbone structure, and then they pack the side-chains associated with the amino acids at each residue. The flexibility of the side-chain makes the search in the space of possible conformations inevitably complex and computationally expensive. The de-novo protein design protocols also rely on similar logical steps: Often an amino acid sequence compatible with a desirable backbone structure is to be inferred (designed)~\citep{dauparas_robust_2022} and then the associated side-chains should be packed to form the full atomic composition of a protein.

Many of the conventional methods for side-chain packing rely on physical models through which they find a rotamer that minimizes a physically-reasoned heuristic energy of the protein fold~\citep{alford_rosetta_2017, krivov_improved_2009, huang_faspr_2020}. However, these computational methods often lack accuracy and speed in their predictions. As deep learning makes strides in protein science, there is a growing effort in developing machine learning methods for rotamer packing. Among these methods is DLPacker~\citep{misiura_dlpacker_2022}, which treats the packing problem as an image transformation. This algorithm characterizes the local environment of a given amino acid backbone within a structure as a 3D image, and uses this model to predict the atomic coordinates of the side chain. It then compares the predicted side-chain to a pre-set library of rotamers to select the closest conformation. AttnPacker~\citep{mcpartlon_end--end_2023}, a more recently developed method, uses a deep graph attention network to model the local geometry of a residue within a structure and is trained to predict the coordinates of the side-chain atoms. Recently, diffusion models over side-chain torsional angles have also being applied, such as DiffPack~\citep{zhang_diffpack_2023}.

Here, we tackle the problem of side-chain packing by learning to directly regress over $\chi$ (torsional) angles, which are main degrees of freedom determining side-chain conformations. We derive and discuss three possible parameterizations of the $\chi$ angles, ultimately settling on regressing over the Sine and Cosine transforms of the angles. We introduce Holographic-Packer (H-Packer), a deep learning method that packs rotamers by first predicting candidate $\chi$ angles from backbone and sequence, and then refines the predictions with a model trained on full-atom structures. Our approach relies on our previously developed holographic convolutions neural network (H-CNN) to characterize amino acid preferences, given their local atomic environment within a structure~\citep{pun_learning_2022}. H-CNN, and by extension H-packer, are locally rotationally (i.e., SO(3)) equivariant, in that they can physically reason about the local geometry of protein structures. Specifically, they  achieve their rotational equivariance by operating fully in the spherical Fourier space.

By directly predicting the side-chain $\chi$-angles, H-packer does not rely on comparing its output with a pre-set library of rotamers, making it computationally more efficient than methods like DLPacker. Furthermore, H-Packer is light-weight (2$\times$3M parameters vs. 208M of AttnPacker) and requires few resources to train (single vs. multiple GPUs for diffusion models like DiffPack). We evaluate the packing performance of H-Packer on standard datasets, and show that it has generally better performance than conventional physics-based methods, and competitive against machine learning solutions. In general, our results suggest that H-Packer has learned complementary features to alternative methods. Our code is freely available at \url{https://github.com/gvisani/hpacker}.

\section{Methods}

In this work, we study the problem of amino-acid side-chain packing using rotationally equivariant neural networks. We introduce H-Packer, a novel yet simple algorithm that predicts side-chain conformations by jointly predicting the values of the key degrees of freedom of a side-chains, i.e., its $\chi$ angles.

\subsection{Modeling side-chain conformations with $\chi$ Angles}

While amino acids are composed of a maximum of 10 heavy atoms (in the case of Tryptophan), their 3D conformations can be uniquely described by the value of at most 4 dihedral angles, referred to as the $\chi$ angles (Figure~\ref{fig:hpacker}A). This reduction in the number of degrees of freedom is granted due to the physical constraints posed on the remaining internal coordinates (bond angles, bond lengths, and dihedral angles - \textit{redundant internal coordinates}). Specifically, the inter-atomic physical interactions within amino acids often constrain these redundant coordinates to a constant, or a well-defined function of the residue's $\chi$ angles. Therefore, predicting $\chi$ angles is the key step in side-chain packing.

H-packer addresses the side-chain packing problem in two steps: (i) it predicts the value of $\chi$ angles, and (ii) it reconstructs the atomic coordinates using the predicted $\chi$ angles and the constrained values of the redundant internal coordinates. Specifically, we evaluate the redundant internal coordinates from a subset of training data (1,700 structures) by leveraging the \texttt{internal\_coords} feature of the \texttt{biopython} package. We empirically verified that the distributions of values were Gaussian with low variance, and resolved to take their \textit{medians} as the ground truth. Substituting these values for the original ones yields a negligible Null Reconstruction error of approximately 0.127\AA\,(Figure~\ref{fig:null_reconstruction}). Notably, this error remains unchanged even when using only 100 reference structures instead of 1,700 (Figure~\ref{fig:null_reconstruction_comparison}).

\subsection{Predicting $\chi$ angles using H-Packer}
We aim to predict the $\chi$ angles associated with a side-chain conformation from the configuration of atoms surrounding a given residue. This atomic neighborhood is associated with the backbone and the side-chain of the neighboring residues in the structure.

\begin{figure}[t]
    \centering
    \resizebox{1.0\textwidth}{!}{
    \includegraphics{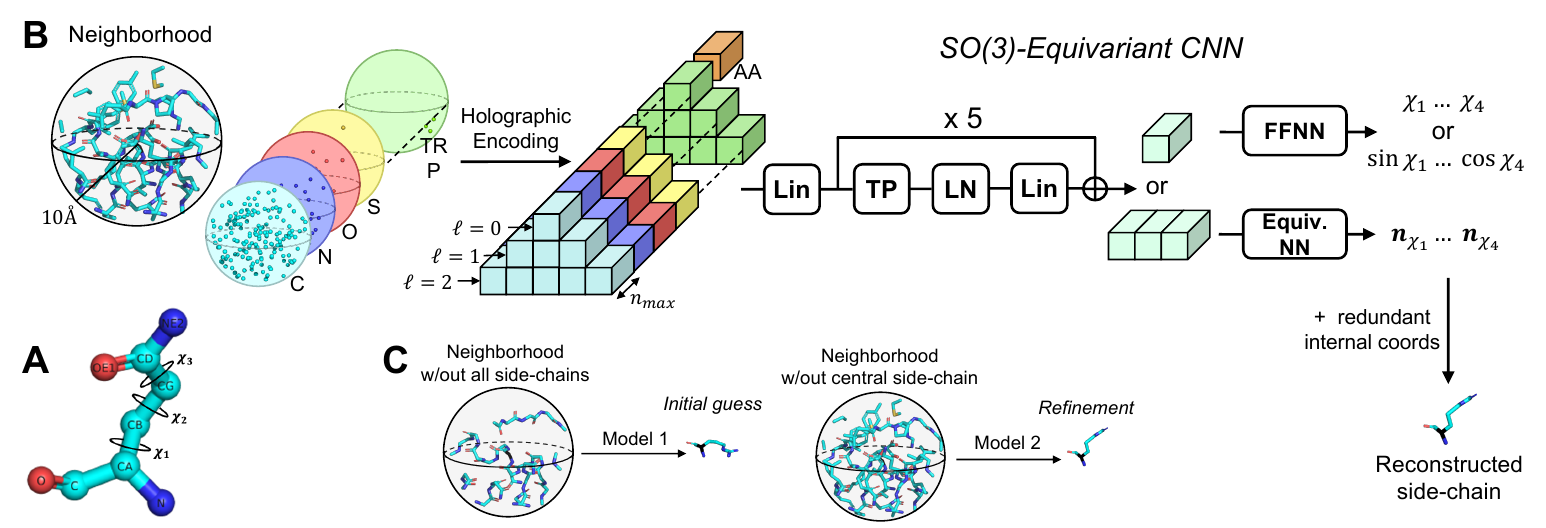}
    }
    \caption{\textbf{Overview of H-Packer.} \textbf{A:} Illustration of Glutamine's $\chi$ angles, of which there are three. \textbf{B:} Schematic shows the H-CNN style network for side-chain packing by first predicting the missing residue's $\chi$ angles from its surrounding atomic environment, and then using the $\chi$ angles to reconstruct the residue's side-chain. As illustrated in \textbf{C}, H-Packer consists of two H-CNN networks, one trained on backbone atoms only and used to make an initial guess, and one trained on full side-chain neighborhoods and used to refine the predictions.}
    \label{fig:hpacker}
\end{figure}

During inference only the coordinates of the backbone atoms are known a priori - alongside the identity of the amino acids they belong to. However, physical interactions with the atoms of other side-chains are the true determinants of a residue's conformation. Therefore, we develop H-Packer into a two-step solution (Figure~\ref{fig:hpacker}C). Specifically, we two train models: one to predict $\chi$ angles from the backbone atoms and amino-acid identity alone, the another to predict $\chi$ angles from full neighborhoods, i.e., by including the true side-chain atoms of the surrounding residues (minus the residue of interest). At inference time, we use the first model to make an \textit{initial guess} of the side-chain conformations, and then a second model to iteratively \textit{refine} the predictions. 

To build the individual models that predict $\chi$ angles, we start by considering their symmetries. Notably, $\chi$ angles are \textit{invariant} to rigid-body transformations (translations and rotations) of the protein (i.e., they are SE(3) invariant). Translation invariance can be satisfied by choosing a well-defined center for a residue of interest; we choose the residue's C-$\alpha$, as it is a common component of all residues and is at the beginning of the side-chain. Then, we still need to take into account rotational invariance about the specified center, which is associated with transformations under the rotation group SO(3).

To respect such rational symmetry, we build SO(3)-\textit{equivariant} models to predict a residue's $\chi$ angles from its surrounding atomic environment. Equivariance is a generalization of invariance whereby when a function's input is transformed by the action of a certain group element (in this case rotation group SO(3)), the output is transformed by the same group element in a well-defined way; equivariant layers ensure both expressivity and efficiency when fitting both invariant and equivariant functions (see Appendix~\ref{sec:mathy_equivariance} for details). To develop these models, we use an approach inspired by our previous work~\cite{visani_holographic-vae_2023, pun_learning_2022}. We consider as input the point cloud of atoms within a radius $r=10\AA$ of the residue's C-$\alpha$ (with or without the neighboring side-chains). To ensure rotational equivariance, we both encode the input in a rotationally equivariant fashion (i.e., a holographic encoding), and use SO(3)-equivariant layers to predict the $\chi$ angles.

\subsubsection{Holographic encoding of the data}
We represent the point clouds of atoms within a structural neighborhood with a density function by summing over (weighted) Dirac-$\delta$ functions,  indicating the presence of atoms at a given position in space: $\rho(r,\theta,\phi) = \sum_{i\in \text{points}} \omega_i \delta(\vr_{i} - \vr)$; here, $\omega_i$ indicates the weight associated with point $i$ at position $\vr_i$. 
We then use 3D Zernike Fourier Transform (ZFT) of the density function to encode the neighborhood into a convenient SO(3) equivariant basis, 

\begin{equation}
    \hat{Z}_{\ell m}^{n} = \sum_{i\in \text{points}} \omega_i R_{n}^{\ell}(r_i) Y_{\ell m}(\theta_i, \varphi_i)
   \label{eqn:spherical_and_radial_with_dirac_ft}
\end{equation}
where $Y_{\ell m}(\theta,\phi) $ is the spherical harmonics of degree $\ell$ and order $m$, and $R^{n}_{\ell}(r)$ is the radial Zernike polynomial in 3D with radial frequency $n \geq 0$ and degree $\ell$. $R^{n}_{\ell}(r)$ is non-zero only for even values of $n-\ell\geq 0$. Notably, the spherical harmonics that describe the angular component of ZFT arise from the irreducible representations of the
3D rotation group SO(3), and form a convenient basis under rotation in 3D (see Appendix~\ref{sec:mathy_equivariance}). Zernike projections in spherical Fourier space can
be understood as a superposition of spherical holograms
of an input point cloud, and thus, we term this operation
as {\em holographic encoding} of the data~\cite{visani_holographic-vae_2023, pun_learning_2022}.

We truncate the Fourier expansion by the  maximum degree $\ell_\text{max}$ and a maximum radial frequency $n_\text{max}$. Additionally, we normalize the Fourier coefficients of each Dirac-$\delta$ function by the sum of the square of its coefficients. We found this normalization to be  beneficial for training, likely due to the avoidance of singularities close to the boundaries. 

Following~\citep{pun_learning_2022} and~\citep{visani_holographic-vae_2023} we incorporate atom-level input features by dividing the holographic encoding into different \textit{channels} (see Figure~\ref{fig:hpacker}). We consider the following two sets: (i) \textbf{Atomic channels}: C, N, O, S, wildcard element excluding hydrogens, partial charge from the Amber99sb force field~\citep{ponder_force_2003}, and (ii)  \textbf{Amino-Acid channels}: one for each of the 20 canonical amino-acids, plus a wildcard channel. We include the charge value in its dedicated channel as the weights $\omega_{i}$ coupled to the point cloud's density function.
While we train the \textit{initial guess} model using both sets of channels (atomic and amino-acid) as input, we only consider the atomic channels for the \textit{refinement} model. We do this in an effort to make the model's predictions more grounded in physical interactions. We condition both models with the identity of the residue of interest by concatenating a linear embedding of its one-hot encoding to the input's invariant ($\ell=0$) features. This is particularly necessary for the refinement model - which is trained only with atomic channels - since it wouldn't otherwise know about the identity of the residue of interest.

\subsubsection{SO(3)-Equivariant neural network architecture}
We use the resulting holograms as inputs to an SO(3)-Equivariant Convolutional Neural Network (Figure~\ref{fig:hpacker}B). The key is to transform the inputs through the network such that all intermediate outputs of the network remain rotationally equivariant. Our resulting model is conceptually divided into three parts:

\textbf{First,} a linear layer that projects data and conditioning to a hidden representation with same number of features per $\ell$.\\
\textbf{Second,} a stack of equivariant blocks connected via additive skip connections, each composed of: (i) feature-wise tensor product nonlinearity, (ii) layer norm with silu nonlinearity, and (iii) a linear layer whose output dimensions are the same as the input's. After the final block, we retain only the features of type $\ell = 0$ or $\ell = 1$ depending on the training objective (Section~\ref{sec:training_objective}). It should be noted that features of type $\ell=0$ are rotationally invariant scalars, whereas those associated with $\ell=1$ are equivariant vectors that transform consistently with the input under rotation. We use $\ell=1$ features to directly learn the orientation of the intersecting planes that define a side-chain's dihedral angles $\chi$ (see Section~\ref{sec:training_objective}).\\
\textbf{Third,} optionally and only for the models with invariant ($\ell = 0$ output), we apply a standard feed-forward neural network with dropout regularization and silu nonlinearity. We refer to Section~\ref{sec:architecture_details} in the appendix for more details on the architecture components.

\subsubsection{Training objectives to infer $\chi$ angles}
\label{sec:training_objective}

We consider three alternative parameterizations of $\chi$ angles, i.e. three possible objective functions:

{\bf  (i) The angle itself.} $\chi$ angles are defined between $-180\degree$ and $180\degree$ with a periodicity such that the angles $-179\degree$ and $179\degree$ are to be considered $2\degree$ apart, not $358\degree$. Thus, plain MSE loss would pose strong and unnatural constraints on the model. To account for this, we mod the predictions to fall in the valid range, and compute the loss between two angles as the minimum between the computed error and $360\degree$ minus the error, resulting in the following loss function:
\EQ
    \mathcal{L}_{\text{angles}}(\{\hat{\chi_i}\}_{i=1}^{N_{\chi}}, \{\chi_i\}_{i=1}^{N_{\chi}}) = \frac{1}{N_\chi} \sum_{i=1}^{N_{\chi}} \min(E_{\chi_i}, \, 2\pi - E_{\chi_i}) \quad \text{where} \quad E_{\chi_i} = (\text{mod}(\hat{\chi_i}, 2\pi) - \chi_i)^2
\EE
where $\hat{\chi_i}$ and $\chi_i$ are  the predicted and the true values of the $i^{th}$ $\chi$, respectively, and $N_\chi$ is the number of $\chi$ angles associated with the residue of interest. In our implementation, the $\chi$ angle domain is scaled and shifted to fall in $[0, 2]$ to make the scale of the loss functions comparable between the three representations of the angles.

{\bf (ii) Sine and Cosine transforms of the angle.} A pair of sine and cosine transformation provides an alternative representation for a $\chi$ angle that accounts for its periodicity and is also rotationally invariant; a similar approach is also considered in concurrent work~\citep{mukhopadhyay_zymepacknet_2023}. We directly predict sine and cosine values by feeding 8 outputs from the network to a tanh activation function, which then form the arguments of  a MSE loss function:
\EQ
\mathcal{L}_{\text{sin-cos}}(\{\hat{\chi_i}\}_{i=1}^{N_{\chi}}, \{\chi_i\}_{i=1}^{N_{\chi}}) = \frac{1}{2 N_\chi} \sum_{i=1}^{N_{\chi}} (\cos \hat{\chi_i}  - \cos \chi_i )^2 +(\sin \hat{\chi_i} - \sin \chi_i)^2
\label{sincosLoss}
\EE
Notably, this loss function is justified by a nice geometric interpretation, whereby it is equivalent to computing the cosine loss between the 2D vectors that describe the $\chi$ angles on the unit circle (proof in Eq.~\ref{eqn:proof_of_equivalence_sin_cos}).

{\bf (iii) Normal vectors to the dihedral plane.} $\chi$ angles are examples of dihedral angles, meaning that they are defined as the angle between two planes. For $\chi$ angles, the two planes are described by subsequent triplets of atoms along the side-chains. Any two subsequent $\chi$ angles share one plane. Therefore, any conformation with $N_{\chi}$ angles can be alternatively described by $N_{\chi}+1$ planes (or their normal vectors); one of these normal vectors is a redundant internal coordinate (defined by backbone + C$\beta$ atoms), while others specify the $N_\chi$ independent degrees of freedom.

We consider training models to predict the dihedral planes' normal vectors: $\vn_{\chi_1}\,...\,\vn_{\chi_4}$. It should be noted that unlike the sine/cosine transformation, the vectors are not invariant to rotations, but \textit{equivariant} of type $\ell = 1$ (geometric vectors) which  can be extracted from the H-Packer equivariant network. We use a cosine loss over the true and predicted vectors:
\EQ
\mathcal{L}_{\text{norms}}(\{\hat{\vn}_{\chi_i}\}_{i=1}^{N_{\chi}}, \{\vn_{\chi_i}\}_{i=1}^{N_{\chi}}) = \frac{1}{N_\chi} \sum_{i=1}^{N_{\chi}} 1 - \langle \hat{\vn}_{\chi_i}, \vn_{\chi_i} \rangle
\label{planeloss}
\EE

{\bf Relevant symmetries in computing loss functions.}
Some amino acid conformations exhibit a rotation symmetry by $\pi$ in some of their $\chi$ angles. For example, $\chi_2$ of Phenylalanine and Tyrosine indicates the torsion of their benzene rings, thus a rotation by $\pi$ leaves the conformation physically unchanged. However, as $\chi$ angles are formally defined by internal atom names, these equivalent conformations are associated with different $\chi$ angle values. We correct for this degeneracy by considering the minimum loss value between considering $\chi$ and $\pi - \chi$ as targets during training and evaluation. When computing the error on the atomic coordinates (generally via Root Mean Square Deviation, RMSD) for the full side-chain, we need to consider other such symmetries between non-$\chi$ atoms, as listed in Table~\ref{table:symmetric_conformations}.

\section{Related Work}
\textbf{Protein side-chain packing.} Methods for side-chain packing can be divided into (older) physics-based algorithms~\citep{alford_rosetta_2017, huang_faspr_2020, krivov_improved_2009, cao_improved_2011, liang_fast_2011} and (newer) machine learning (ML) approaches~\citep{misiura_dlpacker_2022, mcpartlon_end--end_2023, zhang_diffpack_2023, mukhopadhyay_zymepacknet_2023, xu_opus-rota4_2022, nagata_sidepro_2012}. Physics-based approaches generally work by minimizing a hand-crafted energy function over the side-chain conformational space, usually with the help of a rotamer (i.e., side-chain conformation) library to discretize and reduce the dimensionality of such space. Popular algorithms include RosettaPacker from the rosetta suite~\citep{alford_rosetta_2017}, FASPR~\citep{huang_faspr_2020}, and SCWRL~\citep{krivov_improved_2009}. Among ML methods, the most related to this work include: DLPacker~\citep{misiura_dlpacker_2022}, which frames the problem as an image-to-image translation (with "channels" analogous to ours) to predict a 3D "image" of the desired rotamer, which is then matched against a rotamer library to return a valid representation; AttnPacker~\citep{mcpartlon_end--end_2023}, which uses a large (~208M) model derived from the SE(3)-Transformer~\citep{fuchs_se3-transformers_2020} to directly predict the coordinates of side-chain atoms from the backbone structure and the amino-acid sequence. The concurrent ZymePackNet~\citep{mukhopadhyay_zymepacknet_2023} (open source code not available) which autoregressively predicts the sine and cosine of $\chi$ angles, using two graph neural networks in a two-step procedure similar to ours; and DiffPack~\citep{zhang_diffpack_2023}, which consists of four expensive diffusion models over each of the $\chi$ angles, autoregressively used together at inference time.

\textbf{Equivariant neural networks for protein structures.} In recent years, great successes has been achieved in structural biology by leveraging the underlying geometric symmetries in modeling protein structure and surface in the form of developing neural networks that are equivariant to the relevant symmetry transformations~\citep{eismann_hierarchical_2021, gainza_deciphering_2020, jing_learning_2021, pun_learning_2022, visani_holographic-vae_2023}. Specifically, a great deal of literature has been devoted to efficiently modeling 3D atomistic systems using neural networks equivariant to euclidean symmetries~\citep{fuchs_se3-transformers_2020, thomas_tensor_2018, batzner_e3-equivariant_2022, musaelian_learning_2023, kondor_clebsch-gordan_2018}. The drawback is that most such methods are computationally expensive due to computing expensive tensor products between all pairs of neighboring atoms (see Section~\ref{sec:architecture_details} and~\citep{thomas_tensor_2018, batzner_e3-equivariant_2022}). Here,  we greatly reduce computational complexity by constructing equivariant representations of a system about a single natural center (the central residue's C-$\alpha$), following an approach originally designed to model spherical images~\citep{kondor_clebsch-gordan_2018}. Applying this approach to residue-level structure modeling has been proven effective in predicting amino-acid propensities in protein structures~\citep{pun_learning_2022}, as well as compactly encoding residue environments in an unsupervised way for downstream tasks~\citep{visani_holographic-vae_2023}.

\section{Experiments}

\begin{figure}
    \centering
    \resizebox{\textwidth}{!}{
    \begin{tabular}{c c}
        \includegraphics[valign=m,width=0.85\textwidth]{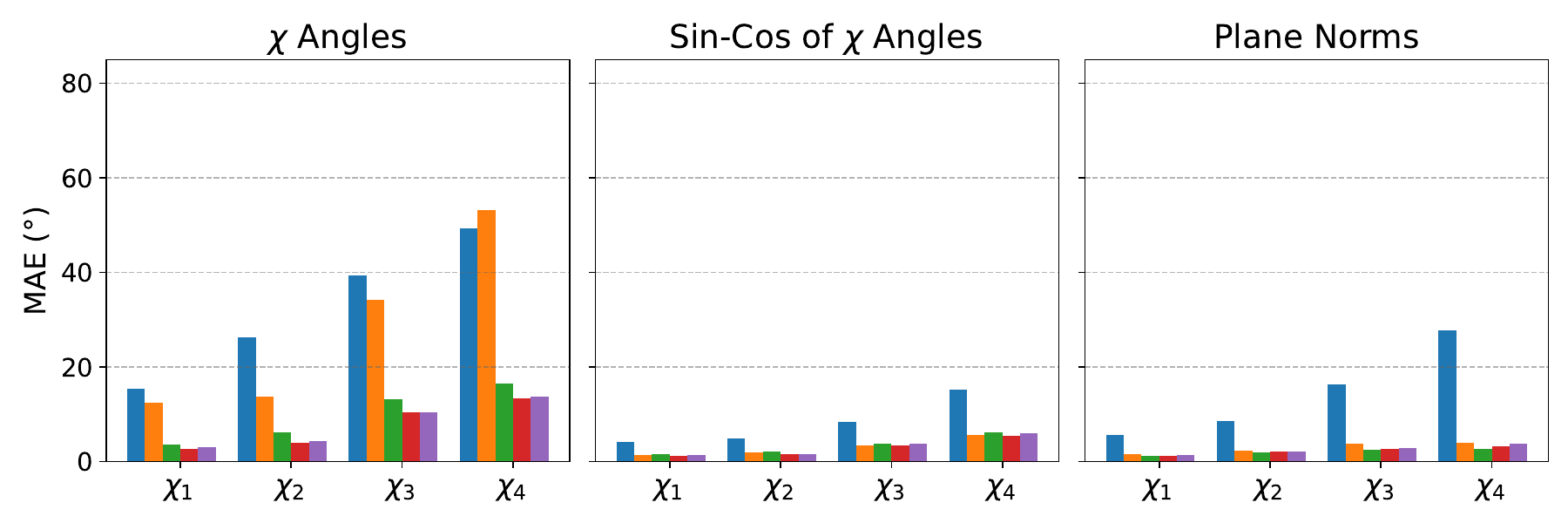}
        &
        \includegraphics[valign=m,width=0.15\textwidth]{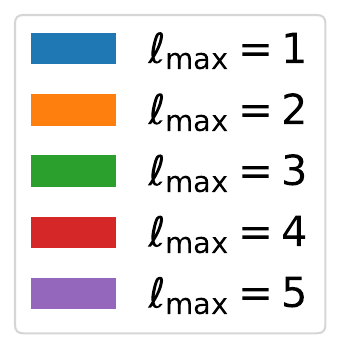}
    \end{tabular}
    }
    \caption{\textbf{Test MAE for the simple task of predicting $\chi$ angles from atomic conformation.} Panels show reconstruction accuracies using three loss functions: the angle $\chi$ itself (left), the sin/cos transform of the angle (center), and the normal vectors to the dihedral planes (right), for different maximum angular degrees $\ell_\text{max}$ (colors). }
    \label{fig:simple_task_performance_by_model_type}
\vspace{-4mm} 
\end{figure}

\subsection{Toy task: inferring $\chi$ angles from atomic coordinates}
\label{sec:simple_task}
We start by studying the behavior of our model on a simple task: predicting (or rather, calculating) $\chi$ angles from the true atomic coordinates of the conformation. We found this to be a useful benchmark to study our model's behavior.

\textbf{Setup.} We randomly select 160 structures from our real task's training set (see below) and split them into 100/30/30 for training/validation/testing, respectively. We collect conformations of all residues presenting $\chi$ angles, and consider only their heavy atoms (C, N, O, S). We then apply the Zernike encoding varying $\ell_{\text{max}}$ from $1$ to $5$ and train models with varying $\ell_{\text{max}}$ consistent with that of the input, as well as with different prediction objectives (angles, sin-cos of angles, plane norms). Crucially, we vary the number of hidden channels (decreasing it with higher $\ell_{\text{max}}$) to keep the number of parameters constant around 330k, and thus, removing differences in model capacity as a contributing factor to performance. We do not condition the models with amino-acid identity to make the problem more challenging, and therefore more interesting. We refer to Section~\ref{sec:training_details} for more details.

\textbf{Results.} Test Mean Absolute Error (MAE) per $\chi$ angle for all models is shown in Figure~\ref{fig:simple_task_performance_by_model_type}, and training curves are shown in the Appendix (Figure~\ref{fig:simple_task_training_curves}). Notably, the Angle model performs the worst, and is unable to recover the true $\chi$ angle with negligible error. The Sin-Cos and Plane Norm models instead recover all $\chi$ angles with very low error (< 5 \AA) with $\ell_{\text{max}} > 1$. 
 It appears that $\ell_{\text{max}}=2$ is the minimum sufficient degree nedded to solve this task with high accuracy. We note that error is higher for later $\chi$ angles. We hypothesise that this is expected for two reasons: (i) later $\chi$ angles depend on atoms that are farther way from the center of the neighborhood, thus having lower angular resolution within the Zernike representation, and (ii) there is simply less training data for them.  Weighting $\chi$ angles in the loss function according to their average frequency partially mitigates the second issue (Figure~\ref{fig:simple_task_performance_by_model_type_with_weighted_loss}). 
 Notably, the fact that the model performs well without explicit knowledge of amino acid identities implies that it can easily infer the amino acid type from the  the number and the relative location of the atoms.

\begin{table}[t]
    \centering
    \resizebox{0.9\textwidth}{!}{
    \begin{tabular}{l c c c c | c c c | c c c}
     \multicolumn{1}{l|}{\bf CASP13} & \multicolumn{4}{c}{Angle MAE $\degree$ $\downarrow$} & \multicolumn{3}{c}{Angle Accuracy \% $\uparrow$} & \multicolumn{3}{c}{Atom RMSD \AA $\downarrow$} \\
     \cline{1-1}
     \textbf{Method} & $\chi_1$ & $\chi_2$ & $\chi_3$ & $\chi_4$ & All & Core & Surface & All & Core & Surface \\
     \hline
     \hline
     SCWRL         & 27.64 & 28.97 & 49.75 & 61.54  &  56.2 & 71.3 & 43.4  &  0.934 & 0.495 & 1.027 \\
     FASPR         & 27.04 & 28.41 & 50.30 & 60.89  &  56.4 & 70.3 & 43.6  &  0.910 & 0.502 & 1.002 \\
     RosettaPacker & 25.88 & 28.25 & 48.13 & 59.82  &  58.6 & 75.3 & 35.7  &  0.872 & 0.422 & 1.001 \\
     DLPacker      & 22.18 & 27.00 & 51.22 & 70.04  &  58.8 & \underline{73.9} & 45.4  &  0.772 & 0.402 & 0.876 \\
     AttnPacker    & \underline{18.92} & \underline{23.17} & \underline{44.89} & 58.98  &  \underline{62.1} & 73.7 & \underline{47.6}  &  \underline{0.669} & \underline{0.366} & \underline{0.775} \\
     DiffPack      & \textbf{15.35} & \textbf{19.19} & \textbf{37.30} & \textbf{50.19}  &  \textbf{69.5} & \textbf{82.7} & \textbf{57.3}  &  \textbf{0.579} & \textbf{0.298} & \textbf{0.696} \\
     \hline
     \hline
     H-Packer$_0^{\ell_{\text{max}}=5}$    & 26.89 & 31.95 & 47.51 & 52.75  &  49.3 & 61.5 & 40.4  &  0.961 & 0.726 & 1.131 \\[1.4pt]
     H-Packer$_2^{\ell_{\text{max}}=5}$     & 23.64 & 29.47 & 45.17 & 53.26  &  54.4 & 69.9 & 43.6  &  0.863 & 0.575 & 1.070 \\[1.4pt]
     H-Packer$_5^{\ell_{\text{max}}=5}$     & 23.60 & 29.40 & \underline{44.91} & \underline{52.91}  &  54.7 & 70.7 & 43.7  &  0.858 & 0.564 & 1.067 \\[1.4pt]
     \hline
     H-Packer$_{up}^{\ell_{\text{max}}=5}$  & \textit{20.03} & \textit{26.88} & \textit{42.74} & \textit{52.07}  &  \textit{58.4} & \textit{75.4} & \textit{46.4}  &  \textit{0.765} & \textit{0.483 }& \textit{0.980} \\
     \hline
    \end{tabular}
    }
    \vspace{6mm}

    \resizebox{0.9\textwidth}{!}{
    \begin{tabular}{l c c c c | c c c | c c c}
     \multicolumn{1}{l|}{\bf CASP14} & \multicolumn{4}{c}{Angle MAE $\degree$ $\downarrow$} & \multicolumn{3}{c}{Angle Accuracy \% $\uparrow$} & \multicolumn{3}{c}{Atom RMSD \AA $\downarrow$} \\
     \cline{1-1}
     \textbf{Method} & $\chi_1$ & $\chi_2$ & $\chi_3$ & $\chi_4$ & All & Core & Surface & All & Core & Surface \\
     \hline
     \hline
     SCWRL         & 33.50 & 33.05 & 51.61 & 55.28  &  45.4 & 62.5 & 33.2  &  1.062 & 0.567 & 1.216 \\
     FASPR         & 33.04 & 32.49 & 50.15 & 54.82  &  46.3 & 62.4 & 34.0  &  1.048 & 0.594 & 1.205 \\
     RosettaPacker & 31.79 & 28.25 & 50.54 & 56.16  &  47.5 & \underline{67.2} & 33.5  &  1.006 & 0.501 & 1.183 \\
     DLPacker      & 29.01 & 33.00 & 53.98 & 72.88  &  48.0 & 66.9 & 33.9  &  0.929 & 0.476 & 1.107 \\
     AttnPacker    & \underline{25.34} & \underline{28.19} & 48.77 & \underline{51.92}  &  \underline{50.9} & 66.2 & 36.3  &  \underline{0.823} & \underline{0.438} & \underline{1.001} \\
     DiffPack      & \textbf{21.91} & \textbf{25.54} & \textbf{44.27} & 55.03  &  \textbf{57.5} & \textbf{77.8} & \textbf{43.5}  &  \textbf{0.770} & \textbf{0.356} & \textbf{0.956} \\
     \hline
     H-Packer$_0^{\ell_{\text{max}}=5}$    & 32.31 & 35.90 & 49.05 & 50.34 & 40.8 & 57.0 & 31.6 & 1.087 & 0.762 & 1.297 \\[1.4pt]
     H-Packer$_2^{\ell_{\text{max}}=5}$    & 29.96 & 34.32 & 47.46 & 50.50 & 45.0 & 65.1 & 34.1 & 1.011 & 0.629 & 1.250 \\[1.4pt]
     H-Packer$_5^{\ell_{\text{max}}=5}$    & 29.61 & 34.03 & \underline{46.72} & \textbf{50.35} & 45.2 & 65.5 & 34.0 & 1.002 & 0.626 & 1.244 \\[1.4pt]
     \hline
     H-Packer$_{up}^{\ell_{\text{max}}=5}$ & \textit{26.58} & \textit{31.54} & \textit{45.67} & \textit{49.46} & \textit{48.1} & \textit{69.5} & \textit{36.2} & \textit{0.915} & \textit{0.534 }& \textit{1.160} \\
     \hline
    \end{tabular}
    }
    \caption{\textbf{Comparative assessment on CASP13 and CASP14.} We present best results in \textbf{bold} and second-best \underline{underlined}. \textit{Italicized} results represent an upper bound to our algorithm's performance. Performance of models other than H-Packer is taken from~\citep{zhang_diffpack_2023}.}
    \label{tab:results_casp13_and_14}
\vspace{-6mm} 
\end{table}

\subsubsection{Side-Chain Packing}

\textbf{Dataset.} We consider the training and validation datasets used in  DLpacker~\citep{misiura_dlpacker_2022}, consisting of 19,436 structures with a maximum inter-protein sequence similarity of 50\%. Unlike DLPacker, we do not remodel structures with PDB-redo~\citep{joosten_pdb_redo_2009} and do not convert selenomethionine residues into methionine. For testing our model, we use the CASP13 and CASP14 targets (82 and 64 structures, respectively). We remove from the training and validation sets any protein that has sequence similarity above 50\% with any of the proteins in the test set.

\textbf{H-Packer training.} We used the Sin/Cos loss function (Eq.~\ref{sincosLoss}) as it was the best-performing loss in our toy-task; while the Plane Norms loss (Eq.~\ref{planeloss}) also performed well in the toy-task, we found that models trained with the Sin/Cos objective were easier to regularize via dropout in the final invariant feed-forward neural network. The {\em initial guess} and the {\em refinement} networks were trained with the same $\ell_{\text{max}}$ of 5 and $n_{\text{max}}=12$; the latter was chosen such that it included at least one radial function with wavelength lower than the minimum interatomic distance. We also considered models trained with $\ell_{\text{max}} = 4$, tuning the number of hidden features to keep the number of trainable parameters the same as the $\ell_{\text{max}} = 5$ models, and equal to $\sim$3M. All models were trained for 10 epochs, keeping the model with lowest validation loss at the end of an epoch; see further details in Section~\ref{sec:training_details}. 
Throughout our experiments, we consider the performance of H-Packer models with different number of rounds of refinement. For example, H-Packer$_0$ denotes the model with {\em no} refinement. For each model, we also compute an {\em upper bound} in performance of the refinement process by tasking the \textit{refinement} model to predict $\chi$ angles from the ground truth neighboring structures (i.e., the toy task); we denote this by H-Packer$_{up}$.

\textbf{Metrics.} In line with previous work~\citep{mcpartlon_end--end_2023, zhang_diffpack_2023}, we evaluate our models on three main metrics. (i) Angle-specific Mean Absolute Error (MAE), (ii) residue-level angle accuracy, defined as the proportion of residues for which the prediction of all $\chi$ angles is within $20\degree$ of the true value, and (iii) average atomic Root Mean Square Deviation (RMSD) of side-chain atoms across residues. We further distinguish between \textit{Surface} and  \textit{Core} residues, as conformations occurring on the surface of proteins are notoriously harder to predict. Surface residues are defined as having at most 15 $\beta$-C within 10~\AA\, of their $\beta$-C, whereas core residues must have at least 20 $\beta$-C's in this range.

\begin{table}[t]
    \centering
    \resizebox{0.9\textwidth}{!}{
    \begin{tabular}{l c c c c | c c c c}
     & \multicolumn{4}{c}{CASP13} & \multicolumn{4}{c}{CASP14} \\
     \textbf{Method} & base & Rec. & Sym. & Rec.+Sym. & base & Rec. & Sym. & Rec.+Sym. \\
     \hline
     \hline
     H-Packer$_0^{\ell_{\text{max}} = 5}$    & 0.961 & 0.943 & 0.923 & 0.906  &  1.087 & 1.070 & 1.050 & 1.034 \\[1.4pt]
     H-Packer$_2^{\ell_{\text{max}} = 5}$    & 0.863 & 0.842 & 0.826 & 0.805  &  1.011 & 0.992 & 0.972 & 0.953 \\[1.4pt]
     H-Packer$_5^{\ell_{\text{max}} = 5}$    & 0.858 & 0.837 & 0.821 & 0.800  &  1.002 & 0.984 & 0.964 & 0.945 \\[1.4pt]
     \hline
     H-Packer$_{up}^{\ell_{\text{max}} = 5}$ & 0.765 & 0.741 & 0.730 & 0.706  &  0.915 & 0.895 & 0.880 & 0.860 \\
     \hline
    \end{tabular}
    }
    \caption{\textbf{Atom RMSD (\AA $\downarrow$) across all residues with different treatments of the true structure.} \textbf{Rec:} reconstructing the true structure with our data-derived redundant internal coordinates. \textbf{Sym:} considering the additional non-natural symmetries used by AttnPacker.}
    \label{tab:rmsd_inflation}
\vspace{-6mm} 
\end{table}

\begin{table}[t]
    \centering
    \resizebox{1.0\textwidth}{!}{
    \begin{tabular}{l l c c c c | c c c | c c c}
     & & \multicolumn{4}{c}{Angle MAE $\degree$ $\downarrow$} & \multicolumn{3}{c}{Angle Accuracy \% $\uparrow$} & \multicolumn{3}{c}{Atom RMSD \AA $\downarrow$} \\
     & \textbf{H-Packer$_5$} & $\chi_1$ & $\chi_2$ & $\chi_3$ & $\chi_4$ & All & Core & Surface & All & Core & Surface \\
     \hline
     \hline
     \multirow{2}{*}{CASP13}
     & $\ell_{\text{max}} = 4$ & 24.27 & 29.76 & 46.32 & \textbf{52.56} & 53.5 & 68.8 & 43.0 & 0.878 & 0.594 & 1.081 \\
     & $\ell_{\text{max}} = 5$ & \textbf{23.60} & \textbf{29.40} & \textbf{44.91} & 52.91 & \textbf{54.7} & \textbf{70.7} & \textbf{43.7} & \textbf{0.858} & \textbf{0.564} & \textbf{1.067} \\
     \hline
     \multirow{2}{*}{CASP14}
     & $\ell_{\text{max}} = 4$ & 30.36 & 34.38 & 48.76 & 50.62 & 43.5 & 64.1 & 32.5 & 1.024 & 0.648 & 1.260 \\
     & $\ell_{\text{max}} = 5$ & \textbf{29.61} & \textbf{34.03} & \textbf{46.72} & \textbf{50.35} & \textbf{45.2} & \textbf{65.5} & \textbf{34.0} & \textbf{1.002} & \textbf{0.626} & \textbf{1.244} \\
     \hline
    \end{tabular}
    }
    \caption{\textbf{Ablation in $\ell_{\text{max}}$.} Metrics for the other H-packer models can be found in Table~\ref{tab:full_ablation_in_lmax}.}
    \label{tab:lmax_ablation}
\vspace{-6mm} 
\end{table}

\textbf{Comparative Evaluation on CASP13 and CASP14 targets.} 
 Table~\ref{tab:results_casp13_and_14} compare H-Packer's performance in side-chain packing with other computational methods~\citep{mcpartlon_end--end_2023, zhang_diffpack_2023, misiura_dlpacker_2022, krivov_improved_2009, alford_rosetta_2017, huang_faspr_2020}. Despite its simplicity, H-Packer$_5$ is competitive against the state-of-the-art at predicting $\chi_3$ and $\chi_4$, but falls behind on $\chi_1$ and $\chi_2$ predictions. This discrepancy indicates that H-Packer has likely learned complementary features to the other models. Moreover, H-Packer mostly outperforms the physics-based computational algorithms~\citep{krivov_improved_2009, alford_rosetta_2017, huang_faspr_2020} and is competitive with DLPacker~\citep{misiura_dlpacker_2022} in all our performance metrics. Interestingly, H-Packer is consistently better than physics-based approaches in terms of overall Atom RMSD, but tends to fall shorter on Angle Accuracy. We present error distrubtions for H-Packer in Figures~\ref{fig:casp13_error_distributions_part_1},~\ref{fig:casp13_error_distributions_part_2},~\ref{fig:casp14_error_distributions_part_1}, and~\ref{fig:casp14_error_distributions_part_2}.

Interestingly, while H-Packer predictions are improved upon using refinement networks, the performance saturates after 2 steps of refinement; the accuracies after 5 iterations of refinement are comparable to those after only 2 steps (Table~\ref{tab:results_casp13_and_14}). Therefore, it is unlikely that further refinement could improve H-Packer's performance to reach its upper bound performance. We hypothesise that training H-Packer to produce confidence scores might help in developing site-specific convergence criteria to help bridge the gap~\citep{mcpartlon_end--end_2023, zhang_diffpack_2023}.

\textbf{Ablation in $\ell_{\text{max}}$.} Table~\ref{tab:lmax_ablation} shows how changing $\ell_{\text{max}}$ (from 4 to 5) impacts  the performance of H-Packer. For the same model capacity, using higher $\ell_{\text{max}}$ consistently yields better performance, indicating that higher angular resolutions of the input can be beneficial for learning this task. This performance improvement comes with a trade-off in  training and inference time, which scale superlinearly with $\ell_{\text{max}}$ unless the Tensor Product computation is adequately constrained~\citep{cobb_efficient_2021}. For reference, training  our models with $\ell_{\text{max}} = 5$ takes $\sim$40\% longer than those with $\ell_{\text{max}} = 4$. We leave the hyperparameter optimization of   $\ell_{\text{max}}$ to future work. 

\textbf{On computing RMSD fairly.} In Table~\ref{tab:results_casp13_and_14} we report RMSD computed by measuring the distance between the coordinates of true and predicted atoms, modulo the symmetries we report in~\ref{table:symmetric_conformations}. However, other algorithms such as AttnPacker~\citep{mcpartlon_end--end_2023} consider other symmetries as well, sometimes even between atoms of differing chemical elements. Though these symmetries reflect spatially similar conformations (such as a flip of the Histidine ring), they result in inflated RMSD scores. We show the effect of this inflation on H-Packer predictions in Table~\ref{tab:rmsd_inflation}. In the same table, we also show the RMSD computed against true structures that have been "reconstructed" using the true $\chi$ angles and the constant values that we use for redundant internal coordinate within H-Packer; we do this in an effort to disentangle the Null Reconstruction Error (Figure~\ref{fig:null_reconstruction}) from the error given by mistakes in $\chi$ angle prediction.

\textbf{Speed.} Table~\ref{tab:speed} shows relative reconstruction speeds for several packing algorithms. Using the current implementation of the reconstruction algorithm, the best-performing H-Packer model is about 7x faster than the popular algorithm RosettaPacker and 6x faster than DLPacker; however, it is considerably slower than AttnPacker. Speed can be considerably cut down by half at the expense of minor performance degradation using two refinement iterations instead of five. However, more considerable speed gains may be achieved by CPU parallelization, when computing  holographic encodings of structural neighborhood during initial data processing. Indeed, each initial guessing and refinement step of H-Packer predicts all $\chi$ angles at once, but in the current implementation holographic encodings are computed in series, creating a bottleneck that currently accounts for ~88\% of the inference time (10\% is atom placement, and only 2\% is making the actual predictions on GPU). We plan on optimizing this aspect in future iterations of the model.

\begin{table}[t]
    \centering
    \resizebox{1.0\textwidth}{!}{
    \begin{tabular}{l c c c c c c c c}
        \textbf{Method} & HPacker$_{5}^{\ell_{\text{max}}=5}$ & HPacker$_{2}^{\ell_{\text{max}}=5}$ & HPacker$_{0}^{\ell_{\text{max}}=5}$ & AttnPack & DLPack & RosPack & FASPR & SCWRL4 \\
        \hline
        \textbf{Rel. Time} & 1.00 & 0.51 & 0.18 & 0.05 & 5.70 & 6.96 & 0.02 & 0.67 \\
        \hline
    \end{tabular}
    }
    \caption{\textbf{Relative times to undertake full atomic reconstruction.} In our current (unoptimized) implementation, HPacker$_{5}^{\ell_{\text{max}}=5}$ takes 1,482s to reconstruct the 82 CASP13 targets on a single NVIDIA A40 GPU. Times for the other methods were taken from ~\citep{mcpartlon_end--end_2023}.}
    \label{tab:speed}
\vspace{-6mm} 
\end{table}

\section{Discussion}
In this paper, we present H-Packer, a novel algorithm for predicting side-chain conformations by jointly regressing over the side-chain's $\chi$ angles. H-packer is composed of two simple and fast rotationally equivariant neural networks, the first one is used for making an \textit{initial guess} using the coordinates of backbone atoms alongside residue identity information, while the second one \textit{refines} the predictions by considering the predicted coordinates of the neighboring side-chain atoms. We carefully study three alternative objective functions, eventually deciding on using a geometrically justified loss function over the sine and cosine of $\chi$ angles. Our experiments show that H-packer is competitive against physics-based methods and some machine-learning solutions, but its performance still lags behind the state-of-the-art at predicting $\chi$ angles closer to the backbone. Overall, the lack of consistent comparative patterns in performance metrics suggests that H-Packer learns features complementary to other approaches. In addition, the formulation of H-packer makes it amenable to easy-to-achieve CPU parallelization to speed up its already fast inference predictions.
We further emphasize that H-Packer is remarkably lightweight - $2\times 3$M parameters vs. 208M of AttnPacker - and requires few resources to train - single GPU at $< 1$ hour per epoch vs. 4 GPUs for 400 epochs for DiffPack (unknown total time).
Limitations of the model include: its inability to distinguish between covalent and non-covalent interactions as atomic interactions are not explicitly encoded into the network, and its inherently lower angular resolution further away from a neighborhood's center. Future areas of improvement include: enhancing angular resolution by scaling up   $\ell_{\text{max}}$ while adjusting the architecture to reduce the resulting computational complexity, and training a confidence model for the predictions and using it to inform the refinement process.

\section{Acknowledgements}
This work has been supported by the National Institutes of Health MIRA award (R35 GM142795), the CAREER award from the National Science Foundation (grant No: 2045054), and the Allen School Computer Science \& Engineering Research Fellowship from the Paul G. Allen School of Computer Science \& Engineering at the University of Washington. This work is also supported, in part, through the Departments of Physics and Computer Science and Engineering, and the College of Arts and Sciences at the University of Washington.


\begin{thebibliography}{10}

\bibitem{dauparas_robust_2022}
J.~Dauparas, I.~Anishchenko, N.~Bennett, H.~Bai, R.~J. Ragotte, L.~F. Milles,
  B.~I.~M. Wicky, A.~Courbet, R.~J. de~Haas, N.~Bethel, P.~J.~Y. Leung, T.~F.
  Huddy, S.~Pellock, D.~Tischer, F.~Chan, B.~Koepnick, H.~Nguyen, A.~Kang,
  B.~Sankaran, A.~K. Bera, N.~P. King, and D.~Baker.
\newblock Robust deep learning–based protein sequence design using
  {ProteinMPNN}.
\newblock {\em Science}, 378(6615):49--56, October 2022.
\newblock Publisher: American Association for the Advancement of Science.

\bibitem{alford_rosetta_2017}
Rebecca~F. Alford, Andrew Leaver-Fay, Jeliazko~R. Jeliazkov, Matthew~J.
  O'Meara, Frank~P. DiMaio, Hahnbeom Park, Maxim~V. Shapovalov, P.~Douglas
  Renfrew, Vikram~K. Mulligan, Kalli Kappel, Jason~W. Labonte, Michael~S.
  Pacella, Richard Bonneau, Philip Bradley, Roland~L. Dunbrack, Rhiju Das,
  David Baker, Brian Kuhlman, Tanja Kortemme, and Jeffrey~J. Gray.
\newblock The {Rosetta} {All}-{Atom} {Energy} {Function} for {Macromolecular}
  {Modeling} and {Design}.
\newblock {\em Journal of Chemical Theory and Computation}, 13(6):3031--3048,
  June 2017.

\bibitem{krivov_improved_2009}
Georgii~G. Krivov, Maxim~V. Shapovalov, and Roland~L. Dunbrack~Jr.
\newblock Improved prediction of protein side-chain conformations with
  {SCWRL4}.
\newblock {\em Proteins: Structure, Function, and Bioinformatics},
  77(4):778--795, 2009.
\newblock \_eprint: https://onlinelibrary.wiley.com/doi/pdf/10.1002/prot.22488.

\bibitem{huang_faspr_2020}
Xiaoqiang Huang, Robin Pearce, and Yang Zhang.
\newblock {FASPR}: an open-source tool for fast and accurate protein side-chain
  packing.
\newblock {\em Bioinformatics (Oxford, England)}, 36(12):3758--3765, June 2020.

\bibitem{misiura_dlpacker_2022}
Mikita Misiura, Raghav Shroff, Ross Thyer, and Anatoly~B. Kolomeisky.
\newblock {DLPacker}: {Deep} learning for prediction of amino acid side chain
  conformations in proteins.
\newblock {\em Proteins: Structure, Function, and Bioinformatics},
  90(6):1278--1290, 2022.
\newblock \_eprint: https://onlinelibrary.wiley.com/doi/pdf/10.1002/prot.26311.

\bibitem{mcpartlon_end--end_2023}
Matthew McPartlon and Jinbo Xu.
\newblock An end-to-end deep learning method for protein side-chain packing and
  inverse folding.
\newblock {\em Proceedings of the National Academy of Sciences},
  120(23):e2216438120, June 2023.
\newblock Publisher: Proceedings of the National Academy of Sciences.

\bibitem{zhang_diffpack_2023}
Yangtian Zhang, Zuobai Zhang, Bozitao Zhong, Sanchit Misra, and Jian Tang.
\newblock {DiffPack}: {A} {Torsional} {Diffusion} {Model} for {Autoregressive}
  {Protein} {Side}-{Chain} {Packing}, June 2023.
\newblock arXiv:2306.01794 [cs, q-bio].

\bibitem{pun_learning_2022}
Michael~N. Pun, Andrew Ivanov, Quinn Bellamy, Zachary Montague, Colin LaMont,
  Philip Bradley, Jakub Otwinowski, and Armita Nourmohammad.
\newblock Learning the shape of protein micro-environments with a holographic
  convolutional neural network, November 2022.
\newblock arXiv:2211.02936 [physics, q-bio].

\bibitem{visani_holographic-vae_2023}
Gian~Marco Visani, Michael~N. Pun, Arman Angaji, and Armita Nourmohammad.
\newblock Holographic-({V}){AE}: an end-to-end {SO}(3)-{Equivariant}
  ({Variational}) {Autoencoder} in {Fourier} {Space}, June 2023.
\newblock arXiv:2209.15567 [physics].

\bibitem{ponder_force_2003}
Jay~W. Ponder and David~A. Case.
\newblock Force fields for protein simulations.
\newblock {\em Advances in Protein Chemistry}, 66:27--85, 2003.

\bibitem{mukhopadhyay_zymepacknet_2023}
Abhishek Mukhopadhyay, Amit Kadan, Benjamin McMaster, J.~Liam McWhirter, and
  Surjit~B. Dixit.
\newblock {ZymePackNet}: rotamer-sampling free graph neural network method for
  protein sidechain prediction, May 2023.
\newblock Pages: 2023.05.05.539648 Section: New Results.

\bibitem{cao_improved_2011}
Yang Cao, Lin Song, Zhichao Miao, Yun Hu, Liqing Tian, and Taijiao Jiang.
\newblock Improved side-chain modeling by coupling clash-detection guided
  iterative search with rotamer relaxation.
\newblock {\em Bioinformatics}, 27(6):785--790, March 2011.

\bibitem{liang_fast_2011}
Shide Liang, Dandan Zheng, Chi Zhang, and Daron~M. Standley.
\newblock Fast and accurate prediction of protein side-chain conformations.
\newblock {\em Bioinformatics}, 27(20):2913--2914, October 2011.

\bibitem{xu_opus-rota4_2022}
Gang Xu, Qinghua Wang, and Jianpeng Ma.
\newblock {OPUS}-{Rota4}: a gradient-based protein side-chain modeling
  framework assisted by deep learning-based predictors.
\newblock {\em Briefings in Bioinformatics}, 23(1):bbab529, January 2022.

\bibitem{nagata_sidepro_2012}
Ken Nagata, Arlo Randall, and Pierre Baldi.
\newblock {SIDEpro}: a novel machine learning approach for the fast and
  accurate prediction of side-chain conformations.
\newblock {\em Proteins}, 80(1):142--153, January 2012.

\bibitem{fuchs_se3-transformers_2020}
Fabian~B. Fuchs, Daniel~E. Worrall, Volker Fischer, and Max Welling.
\newblock {SE}(3)-{Transformers}: {3D} {Roto}-{Translation} {Equivariant}
  {Attention} {Networks}, November 2020.
\newblock arXiv:2006.10503 [cs, stat].

\bibitem{eismann_hierarchical_2021}
Stephan Eismann, Raphael J.~L. Townshend, Nathaniel Thomas, Milind Jagota,
  Bowen Jing, and Ron~O. Dror.
\newblock Hierarchical, rotation-equivariant neural networks to select
  structural models of protein complexes.
\newblock {\em Proteins: Structure, Function, and Bioinformatics},
  89(5):493--501, May 2021.
\newblock arXiv:2006.09275 [cs, q-bio, stat].

\bibitem{gainza_deciphering_2020}
P.~Gainza, F.~Sverrisson, F.~Monti, E.~Rodolà, D.~Boscaini, M.~M. Bronstein,
  and B.~E. Correia.
\newblock Deciphering interaction fingerprints from protein molecular surfaces
  using geometric deep learning.
\newblock {\em Nature Methods}, 17(2):184--192, February 2020.
\newblock Number: 2 Publisher: Nature Publishing Group.

\bibitem{jing_learning_2021}
Bowen Jing, Stephan Eismann, Patricia Suriana, Raphael J.~L. Townshend, and Ron
  Dror.
\newblock Learning from {Protein} {Structure} with {Geometric} {Vector}
  {Perceptrons}, May 2021.
\newblock arXiv:2009.01411 [cs, q-bio, stat].

\bibitem{thomas_tensor_2018}
Nathaniel Thomas, Tess Smidt, Steven Kearnes, Lusann Yang, Li~Li, Kai Kohlhoff,
  and Patrick Riley.
\newblock Tensor field networks: {Rotation}- and translation-equivariant neural
  networks for {3D} point clouds, May 2018.
\newblock arXiv:1802.08219 [cs].

\bibitem{batzner_e3-equivariant_2022}
Simon Batzner, Albert Musaelian, Lixin Sun, Mario Geiger, Jonathan~P. Mailoa,
  Mordechai Kornbluth, Nicola Molinari, Tess~E. Smidt, and Boris Kozinsky.
\newblock E(3)-equivariant graph neural networks for data-efficient and
  accurate interatomic potentials.
\newblock {\em Nature Communications}, 13(1):2453, May 2022.
\newblock Number: 1 Publisher: Nature Publishing Group.

\bibitem{musaelian_learning_2023}
Albert Musaelian, Simon Batzner, Anders Johansson, Lixin Sun, Cameron~J. Owen,
  Mordechai Kornbluth, and Boris Kozinsky.
\newblock Learning local equivariant representations for large-scale atomistic
  dynamics.
\newblock {\em Nature Communications}, 14(1):579, February 2023.
\newblock Number: 1 Publisher: Nature Publishing Group.

\bibitem{kondor_clebsch-gordan_2018}
Risi Kondor, Zhen Lin, and Shubhendu Trivedi.
\newblock Clebsch-{Gordan} {Nets}: a {Fully} {Fourier} {Space} {Spherical}
  {Convolutional} {Neural} {Network}, November 2018.
\newblock arXiv:1806.09231 [cs, stat].

\bibitem{joosten_pdb_redo_2009}
R.~P. Joosten, J.~Salzemann, V.~Bloch, H.~Stockinger, A.-C. Berglund,
  C.~Blanchet, E.~Bongcam-Rudloff, C.~Combet, A.~L. Da~Costa, G.~Deleage,
  M.~Diarena, R.~Fabbretti, G.~Fettahi, V.~Flegel, A.~Gisel, V.~Kasam,
  T.~Kervinen, E.~Korpelainen, K.~Mattila, M.~Pagni, M.~Reichstadt, V.~Breton,
  I.~J. Tickle, and G.~Vriend.
\newblock {PDB}\_redo: automated re-refinement of {X}-ray structure models in
  the {PDB}.
\newblock {\em Journal of Applied Crystallography}, 42(3):376--384, June 2009.
\newblock Publisher: International Union of Crystallography.

\bibitem{cobb_efficient_2021}
Oliver~J. Cobb, Christopher G.~R. Wallis, Augustine~N. Mavor-Parker, Augustin
  Marignier, Matthew~A. Price, Mayeul d'Avezac, and Jason~D. McEwen.
\newblock Efficient {Generalized} {Spherical} {CNNs}, March 2021.
\newblock arXiv:2010.11661 [astro-ph].

\bibitem{bronstein_geometric_2021}
Michael~M. Bronstein, Joan Bruna, Taco Cohen, and Petar Veličković.
\newblock Geometric {Deep} {Learning}: {Grids}, {Groups}, {Graphs},
  {Geodesics}, and {Gauges}, May 2021.
\newblock arXiv:2104.13478 [cs, stat].

\bibitem{geiger_e3nn_2022}
Mario Geiger and Tess Smidt.
\newblock e3nn: {Euclidean} {Neural} {Networks}, July 2022.
\newblock arXiv:2207.09453 [cs].

\bibitem{tung_group_1985}
Wu-Ki Tung.
\newblock {\em Group {Theory} in {Physics}}.
\newblock 1985.

\bibitem{ba_layer_2016}
Jimmy~Lei Ba, Jamie~Ryan Kiros, and Geoffrey~E. Hinton.
\newblock Layer {Normalization}, July 2016.
\newblock arXiv:1607.06450 [cs, stat].

\bibitem{worrall_harmonic_2017}
Daniel~E. Worrall, Stephan~J. Garbin, Daniyar Turmukhambetov, and Gabriel~J.
  Brostow.
\newblock Harmonic {Networks}: {Deep} {Translation} and {Rotation}
  {Equivariance}, April 2017.
\newblock arXiv:1612.04642 [cs, stat].

\bibitem{noauthor_se3-transformers_nodate}
{SE}(3)-{Transformers} for {PyTorch} {\textbar} {NVIDIA} {NGC}.

\bibitem{kingma_adam_2017}
Diederik~P. Kingma and Jimmy Ba.
\newblock Adam: {A} {Method} for {Stochastic} {Optimization}, January 2017.
\newblock arXiv:1412.6980 [cs].

\end{thebibliography}

\newpage

\appendix
\counterwithin{table}{section}
\setcounter{table}{0}
\counterwithin{figure}{section}
\setcounter{figure}{0}
\counterwithin{equation}{section}
\setcounter{equation}{0}

\section{Appendix}

\subsection{More rigorous mathematical background on SO(3)-Equivariance}
\label{sec:mathy_equivariance}

\textbf{Group Invariance and Equivariance.} Intuitively, a function is said to be \textit{invariant} to a certain group of transformations (e.g. 3D rotations) if applying one such transformation to the function's input does not change its output. \textit{Equivariance} is a generalization of invariance whereby when the input is transformed by the action of a certain group element (or rather by a matrix representation parameterized by the group element) the output of the function is transformed by the same group element (i.e., by a matrix representation parameterized by the same group element, but that can be different from the input's representation). In short, equivariant functions transform the input in the same way regardless of its coordinate frame, but do not necessarily discard the coordinate frame information, whereas invariant functions also do the latter. Both of these concepts can be extended to \textit{properties} as well, e.g. ``the mass of a molecule remains constant (is invariant) when rotating it, whereas its dipole moment rotates alongside it (is equivariant)". More formally, a function between two vector spaces $f: X \rightarrow Y$ is said to be equivariant to a group of transformations $\mathfrak{G}$ iff applying any group transformation to the input space of $f$ corresponds to applying the same transformation to the output space (i.e., via a \textit{representation} parametrized by the same group element). Formally: $f(\mD_{X}(\mathfrak{g})\vx) = \mD_{Y}(\mathfrak{g})f(\vx) \,, \forall \vx \in X \land \forall \mathfrak{g} \in \mathfrak{G}$. The group acts on the input and output vector spaces with space-specific representations that are appropriate for the space (i.e., $\mD_{X}$ and $\mD_{Y}$). A group may have different representations, and a special one is the one that always maps to the identity: $\mD_{Y}(\mathfrak{g})=1, \forall \mathfrak{g} \in \mathfrak{G}$; a function on whose output space $\mathfrak{G}$ acts with the identity representation is said to be \textit{invariant} to $\mathfrak{G}$.
In the context of machine learning, building models for which the output is provably invariant/equivariant to the same groups as the target function can avoid expensive data augmentation. However, even when fitting invariant functions, using equivariant layers is advisable - if not necessary~\citep{bronstein_geometric_2021}.

\textbf{Irreducible representations.} How are equivariant layers generally achieved? The key is to look at the group's \textit{irreducible representations} (irreps). These are the group's \textit{smallest} representations, so that any possible representation can be provably decomposed into a direct sum of irreps. Therefore, the group's irreps can be used to describe how the group elements act on any vector space. We can use this fact to build group-equivariant functions by ensuring that both the input and output of the function are composed (via direct sum i.e., concatenation) of features that transform under the group's action under the group's irreps.

\textbf{SO(3)-Equivariance.} The above is often easier said than done, but it has been worked out for SO(3), which is a group describing 3D rotations about a fixed point~\citep{geiger_e3nn_2022, kondor_clebsch-gordan_2018, thomas_tensor_2018}. Spherical Fourier space can be used to conveniently define equivariant  transformation for rotations. For rotations  about a given reference point, the points in 3D can be  expressed  by the resulting spherical coordinates $(r,\theta,\phi)$ about the set origin. Since the radius $r$ (i.e., the distance of a point from to the reference) does not change under rotations about the origin, we will ignore the radial component for now and consider a signal over the sphere of radius $r$, $f(\theta,\phi):~{S}^2(r) \rightarrow \mathbb{R}$. The Fourier transform $\hat{F}$ of the signal on the sphere follows, 
\EQ
	\hat{F}_{\ell m}=\int_0^{2\pi}  \int_0^\pi f(\theta,\phi) Y_{\ell m} (\theta,\phi) \sin \theta \,d\theta\,d\phi
	\label{eq:spherical_ft}
\EE
where $Y_{\ell m}(\theta,\phi)$ is the spherical harmonic of degree $\ell$ and order $m$ defined as
\EQ
	Y_{\ell m}(\theta,\phi) = \sqrt{\frac{2n+1}{4\pi}\frac{(n-m)!}{(n+m)!}}e^{im\phi}P^m_\ell(\cos\theta)
	\label{eq:YLM}
\EE
where $\ell$ is a non-negative integer ($0\leq \ell$), and $m$ is an integer within the interval $-\ell\leq m \leq \ell$. $P^m_\ell(\cos \theta)$ is the Legendre polynomial of degree $\ell$ and order $m$.
The operators that describe how spherical harmonics transform under rotations are called the Wigner D-matrices, denoted by  $D^{\ell}_{mm'}(R)$~\citep{tung_group_1985}.
\EQ
	Y_{\ell m}(\theta,\phi) \overset{\text{rotation:\,} R}{\longrightarrow}\sum_{m' = -\ell}^\ell D^{\ell}_{m'm}(R) Y_{\ell m'}(\theta,\phi)
	\label{eq:YLMWignerD}
\EE
Indeed, Wigner-D matrices are the irreps of SO(3).  Therefore, any vector space that "3D-rotates" can be decomposed into a direct sum of type-$\ell$ features that transform according to the irrep of type $\ell$. For example, features of type $0$ are invariant to rotation (e.g. atomic mass), while features of type $1$ transform as geometric vectors (e.g. dipole moment). Thus, to build an SO(3)-equivariant model we start by projecting the data onto a convenient SO(3)-equivariant basis via the spherical harmonics. We then leverage a suite of rules that allows one to build learnable layers without breaking equivariance, and transform the input into a new representation composed of features within the same range of possible types. One key transformation rule is the Clebsch-Gordan Tensor Product~\citep{tung_group_1985} that is commonly used to inject nonlinearity (to be precise bi-linearity) in the SO(3) equivariant neural networks~\citep{kondor_clebsch-gordan_2018}.

\subsection{Equivariant architecture components}
\label{sec:architecture_details}

\textbf{Linearity.} The equivariant linear layer consists of a set of linear projections each acting on the set of features sharing the same type $\ell$. It practically equates to $\ell_{\text{max}} + 1$ standard linear layers with no bias except for the $\ell=0$ case, and with the consideration that all $2\ell+1$ moments of the same feature are processed together. Formally, let $\vh_{\ell} \in \mathbb{R}^{C \times (2\ell + 1)}$ be a set of $C$ features of type $\ell$. Then, we learn weight matrix $\mW_{\ell} \in \mathbb{R}^{C \times K}$ that linearly maps $h_{\ell}$ to $\overline{h}_{\ell} \in \mathbb{R}^{K \times (2\ell + 1)}$; if $\ell=0$, $\mathbf{b}_{\ell} \in\mathbb{R}^{K}$ is also learned:
\begin{equation}
    \overline{\vh}_{\ell} = \mW_{\ell}^{T} \vh_{\ell} \,\,(+ \,\mathbf{b}_{\ell})
    \label{eqn:linearity}
\end{equation}

\textbf{Tensor Product Nonlinearity.} Arguably the most important SO(3)-Equivariant operation is the Clebsch-Gordan (CG) Tensor Product. It is the only known operation capable of nonlinearly coupling (i.e. mixing information between) features of different type $\ell$ and with different momentum index $m$. The CG tensor product combines two features of degrees $\ell_1$ and $\ell_2$ to produce another feature of degree $|\ell_2-\ell_1|\leq\ell_3\leq |\ell_1+\ell_2|$. Let $\vh_{\ell} \in \mathbb{R}^{2\ell + 1}$ be a generic degree $\ell$ tensor, with individual components $\hat h_{\ell m}$ for $-\ell\leq m \leq \ell $. The CG tensor product is given by,
\begin{equation}
\begin{aligned}
    \hat  h_{\ell_3 m_3} &= (\vh_{\ell_1} \otimes_{cg} \vh_{\ell_2})_{\ell_3 m_3} \\
    &= \sum_{m_1 = -\ell_1}^{\ell_1} \sum_{m_2 = -\ell_2}^{\ell_2} C_{(\ell_1 m_1)(\ell_2 m_2)}^{(\ell_3 m_3)} \hat h_{\ell_1 m_1} \hat h_{\ell_2 m_2}
    \label{eqn:cg_tensor_product}
\end{aligned}
\end{equation}
where $C_{(\ell_1 m_1)(\ell_2 m_2)}^{(\ell_3 m_3)}$ are the Clebsch-Gordan coefficients. Similar to spherical harmonics, Clebsch-Gordan tensor products also appear in quantum mechanics, and they are used to express couplings between angular momenta.

Here, we use the CG Tensor Product as the primary nonlinear activation of our networks, as originally prescribed by~\citep{kondor_clebsch-gordan_2018}, and crucially the only operation that can transfer information across features of different types $\ell$. Following~\citep{cobb_efficient_2021} and~\citep{visani_holographic-vae_2023}, we compute the Tensor Product feature-wise (i.e. we do not compute it across features with a different index) to significantly decrease computation time. We refer the reader to~\citep{cobb_efficient_2021} and~\citep{visani_holographic-vae_2023} for details. 

\textbf{Layer Norm Nonlinearity.} This consists of applying a standard Layer Norm~\citep{ba_layer_2016} to the \textit{norms} of type-$\ell$ features (which are invariant), and then feeding the normalized norm into a standard nonlinear activation function. This layer effectively combines the equivariant layer norm in \texttt{e3nn}~\citep{geiger_e3nn_2022} with the Norm Nonlinearity originally used in Harmonic Networks~\cite{worrall_harmonic_2017}, and then adapted to the SO(3) domain by Tensor Field Networks~\citep{thomas_tensor_2018}. It is also used in the SE(3)-Transformer ~\citep{fuchs_se3-transformers_2020, noauthor_se3-transformers_nodate}.

The SO(3)-equivariant architecture components were implemented with the help of \texttt{e3nn} primitives~\citep{geiger_e3nn_2022}. We emphasize that we do not ablate all the architectural components, and different choices would be possible. Specifically, we do not ablate over the choice of normalization and of invariant nonlinearity function: in preliminary experiments, other options seemed to perform equally. Other components are necessary or were otherwise found to be useful. For example the tensor product is necessary to ensure that information flows across different $\ell$s, and tuning dropout was found to be greatly useful to prevent overfitting.

\subsection{Proof of equivalence between MSE and cosine loss in predicting Sine and Cosine of $\chi$ angles}
Let ${\chi_i}$ and $\hat\chi_i$ be the true and the predicted angle values for a residue's $i^{th}$ $\chi$ angle, respectively. Let ${{\bf v}_i\equiv~[\sin {\chi_i},\cos {\chi_i}]}$ and ${\hat{\bf v}_i\equiv[\sin { \hat \chi_i},\cos  {\hat \chi_i}]}$ denote their respective 2D vectors on the unit circle, whose two components are the sine and cosine of the angles themselves. Then we have:
\begin{align}
 \nonumber    \text{MSE}_{\text{sin-cos}}(\hat\chi_i, \chi_i)
    \nonumber                                &= \frac{1}{2}(\cos \hat \chi_i - \cos \chi_i)^2 + \frac{1}{2}(\sin\hat \chi_i - \sin \chi_i)^2 \\
             \nonumber                       &= 1 - \cos \hat \chi_i  \cos {\chi_i} - \sin \hat \chi_i  \sin {\chi_i} \\
               \nonumber                     &= 1 - \langle \hat{\bf v}_{i},{\bf v}_i \rangle \\
                                    &= \text{CosineLoss}(\hat{\bf v}_{i}, {\bf v}_i)
    \label{eqn:proof_of_equivalence_sin_cos}
\end{align}
where the jump from step 1 to step 2 is granted by the trigonometric identity $\sin^2 \theta + \cos^2 \theta = 1$.

\subsection{Training Details}
\label{sec:training_details}

\subsubsection{Toy Task}
All  models were constructed with 5 equivariant blocks and no invariant feed-forward neural network (FFNN) for the two models predicting invariant quantities. All models were trained using Adam~\citep{kingma_adam_2017} for 30 epochs with learning rate 0.001 and a batch size of 32. The model exhibiting lowest validation loss was used.

\subsubsection{Side-chain Packing}
Models with $\ell_{\text{max}} = 4$ have five equivariant blocks with a per-$\ell$ hidden feature side of 128. For models with $\ell_{\text{max}} = 5$ we use 5 blocks with per-$\ell$ size 96. In doing so, the two models have comparable number of parameters (3M). All models have a 3-layer FFNN with silu nonlinearity and dropout normalization rate of 0.1. We found it useful to tune he dropout rate to prevent overfitting. We train all models for 10 epochs, keeping the model with lowest validation loss at the end of an epoch (convergence usually happened by epoch $\sim$8); models with $\ell_{\text{max}} = 5$ took roughly 50 minutes per epoch to train on a single NVIDIA A40 GPU, while models with $\ell_{\text{max}} = 4$ took 35 minutes.

\begin{figure}
    \centering
    \includegraphics[width=0.9\textwidth]{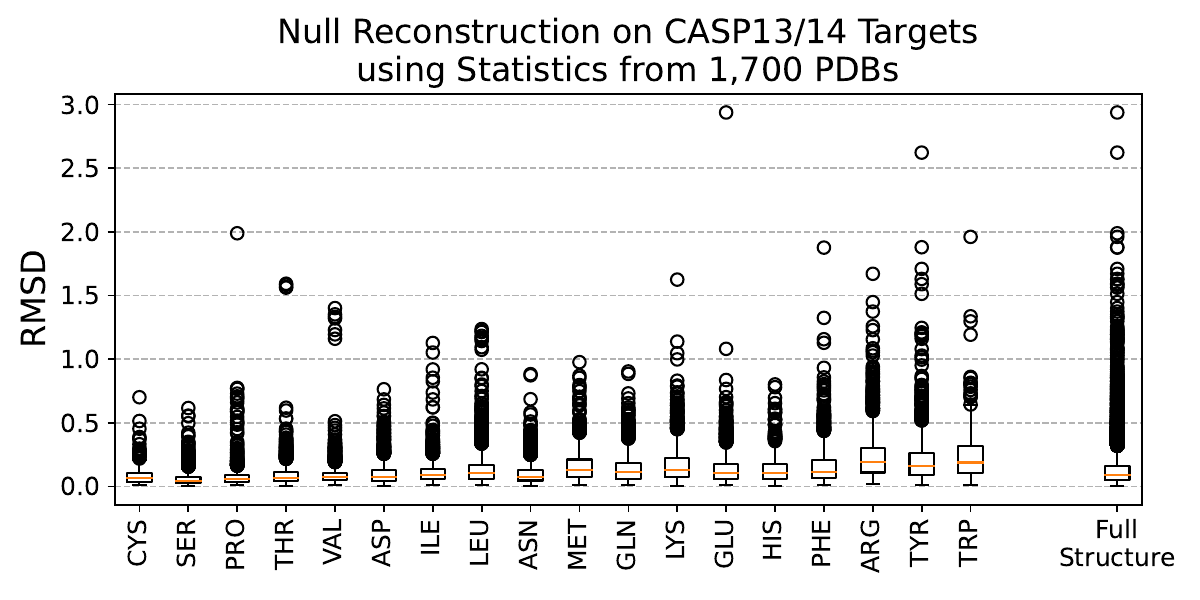}
    \caption{\textbf{Null Reconstruction Error on Test data by using data-derived values as redundant internal coordinates.} Effectively, each redundant internal coordinate (i.e. excluding chi angles) gets substituted from a single value computed as the median of the corresponding value in a reference dataset. Across all structures, the average Null Reconstruction RMSD is 0.127 \AA. We notice a small number of outliers (single residues with abnormally high RMSD), but we do not investigate the causes.}
    \label{fig:null_reconstruction}
\end{figure}

\begin{figure}
    \centering
    \includegraphics[width=0.9\textwidth]{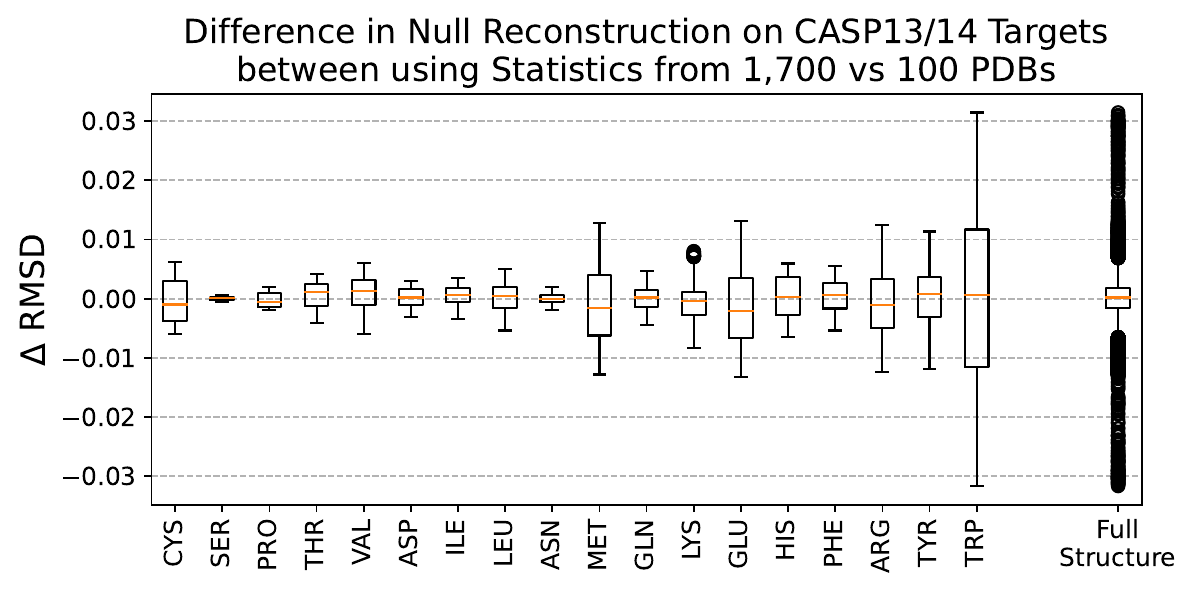}
    \caption{\textbf{Difference in Null Reconstruction Error on Test data by using a smaller set of structures for reference.}}
    \label{fig:null_reconstruction_comparison}
\end{figure}

\begin{table}
\centering
    \caption{\textbf{List of conformational symmetries that we take into consideration.}} 
    \label{table:symmetric_conformations}
    \begin{tabular}{c | c | c}
        \textbf{Amino-Acid} & \textbf{$\mathbf{\pi}$-Symmetric $\mathbf{\chi}$} & \textbf{Pairs of Equivalent Atom names} \\
        \hline
        ARG & - & NH1-NH2 \\
        TYR & $\chi_2$ & CD1-CE1, CD2-CE2 \\
        PHE & $\chi_2$ & CD1-CE1, CD2-CE2 \\
        ASP & $\chi_2$ & OD1-OD2 \\
        GLU & $\chi_3$ & OE1-OE2
    \end{tabular}
\end{table}

\begin{figure}
    \centering
    \begin{tabular}{c}
        \includegraphics[valign=m,width=1.0\textwidth]{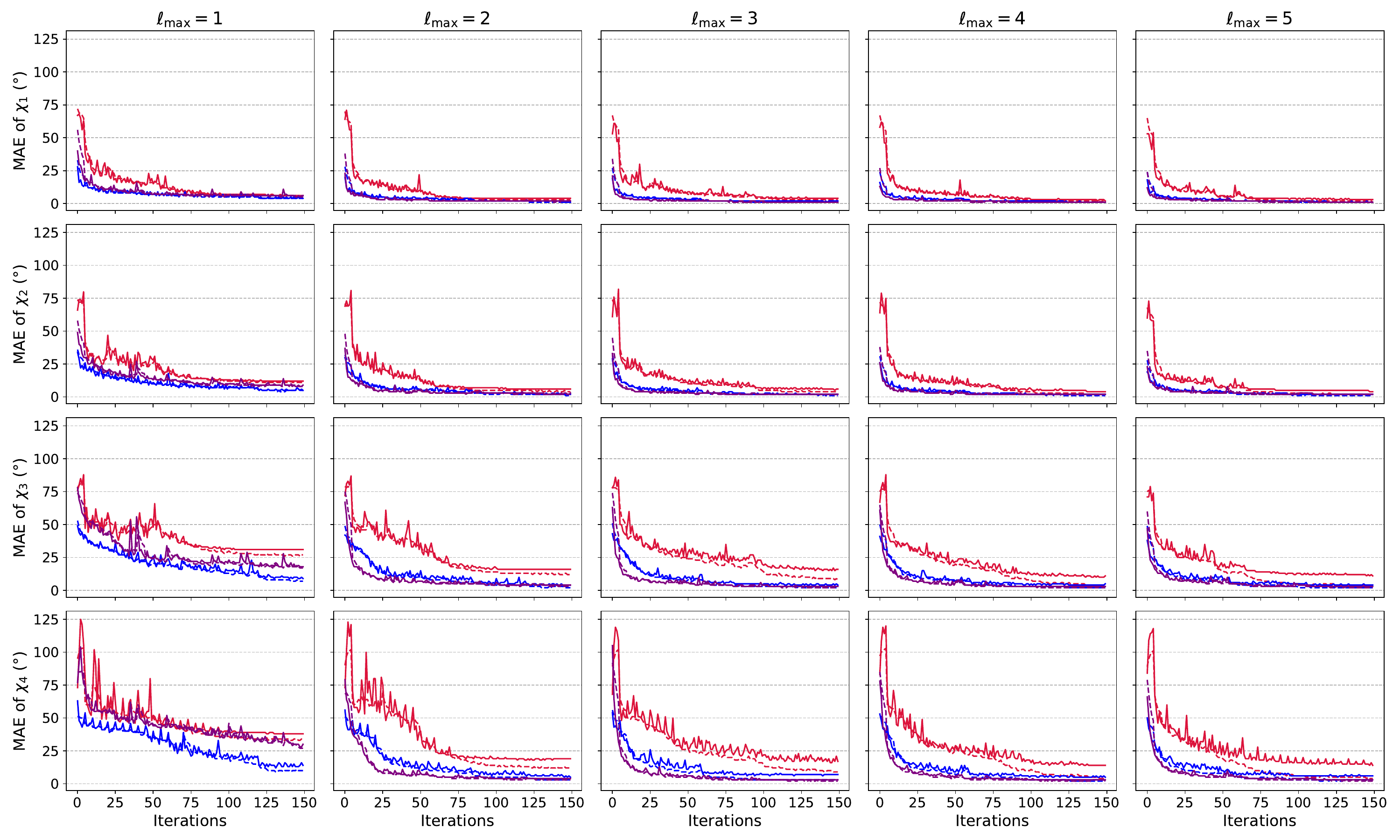}
        \\
        \includegraphics[valign=m,width=0.20\textwidth]{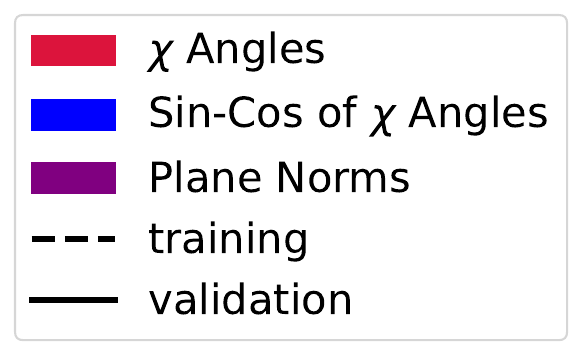}
    \end{tabular}
    \caption{\textbf{$\chi$ angles error trace during training for the simple task of predicting $\chi$ angles from true amino-acid conformations.} The Sin-Cos and Plane Norms models show comparable convergence curves, except for $\ell_{\text{max}} = 1$ where the Plane Norms model struggles with $\chi_3$ and $\chi_4$. The Angles model has the worst convergence trace, even overfitting for high $\ell_{\text{max}}$.}
    \label{fig:simple_task_training_curves}
\end{figure}

\begin{figure}
    \centering
    \resizebox{\textwidth}{!}{
    \begin{tabular}{c c}
        \includegraphics[valign=m,width=0.85\textwidth]{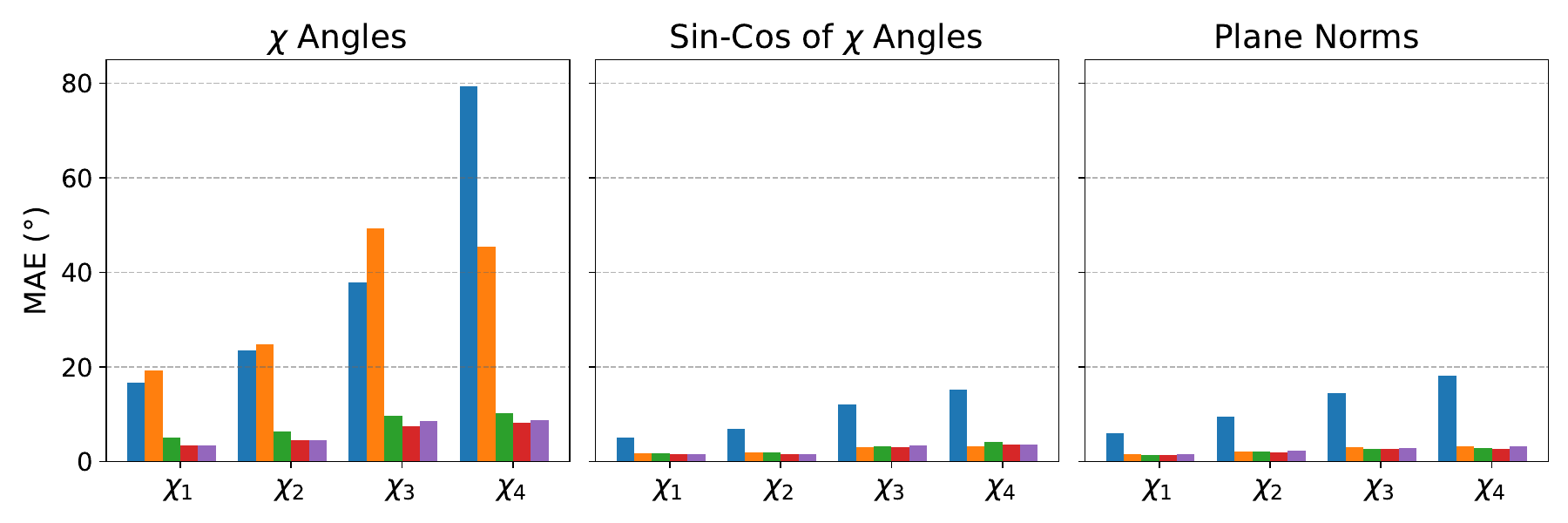}
        &
        \includegraphics[valign=m,width=0.15\textwidth]{lmax_legend.pdf}
    \end{tabular}
    }
    \caption{\textbf{Test MAE for the simple task of predicting $\chi$ angles from atomic conformation, with loss weighted by frequency.} Models trained with weighting $\chi$ angles in the loss function based on their frequency. We use the following weights for $\chi_{1}$ to $\chi_{4}$, which are derived from how many residues have which $\chi$ angle: $[1.0, 18.0/14.0, 18.0/5.0, 18.0/2.0]$. We then re-scale them to make sure the absolute scale of the loss function is the same as it was in the un-weighted case.}
    \label{fig:simple_task_performance_by_model_type_with_weighted_loss}
\end{figure}

\begin{table}[t]
    \centering
    \resizebox{1.0\textwidth}{!}{
    \begin{tabular}{l l c c c c | c c c | c c c}
     & & \multicolumn{4}{c}{Angle MAE $\degree$ $\downarrow$} & \multicolumn{3}{c}{Angle Accuracy \% $\uparrow$} & \multicolumn{3}{c}{Atom RMSD \AA $\downarrow$} \\
     & \textbf{Method} & $\chi_1$ & $\chi_2$ & $\chi_3$ & $\chi_4$ & All & Core & Surface & All & Core & Surface \\
     \hline
     \hline
     \multirow{8}{*}{CASP13}
     & H-Packer$_0$ $\ell_{\text{max}} = 4$    & 27.60 & 32.31 & 47.87 & 52.77 & 48.3 & 59.1 & \textbf{40.5} & 0.979 & 0.771 & 1.136 \\
     & H-Packer$_0$ $\ell_{\text{max}} = 5$    & \textbf{26.89} & \textbf{31.95} & \textbf{47.51 }& \textbf{52.75} & \textbf{49.3} & \textbf{61.5} & 40.4 & \textbf{0.961} & \textbf{0.726} & \textbf{1.131} \\
     \cline{2-12}
     & H-Packer$_2$ $\ell_{\text{max}} = 4$    & 24.36 & 29.94 & 46.01 & \textbf{52.89} & 53.0 & 67.7 & 42.5 & 0.884 & 0.610 & 1.083 \\
     & H-Packer$_2$ $\ell_{\text{max}} = 5$    & \textbf{23.64} & \textbf{29.47} & \textbf{45.17} & 53.26 & \textbf{54.4} & \textbf{69.9} & \textbf{43.6} & \textbf{0.863} & \textbf{0.575} & \textbf{1.070} \\
     \cline{2-12}
     & H-Packer$_5$ $\ell_{\text{max}} = 4$    & 24.27 & 29.76 & 46.32 & \textbf{52.56} & 53.5 & 68.8 & 43.0 & 0.878 & 0.594 & 1.081 \\
     & H-Packer$_5$ $\ell_{\text{max}} = 5$    & \textbf{23.60} & \textbf{29.40} & \textbf{44.91} & 52.91 & \textbf{54.7} & \textbf{70.7} & \textbf{43.7} & \textbf{0.858} & \textbf{0.564} & \textbf{1.067} \\
     \cline{2-12}
     & H-Packer$_{up}$ $\ell_{\text{max}} = 4$ & 20.78 & 27.48 & 44.58 & \textbf{51.97} & 56.8 & 72.7 & 45.6 & 0.790 & 0.514 & 0.999 \\
     & H-Packer$_{up}$ $\ell_{\text{max}} = 5$ & \textbf{20.03} & \textbf{26.88} & \textbf{42.74} & 52.07 & \textbf{58.4} & \textbf{75.4 }& \textbf{46.4} & \textbf{0.765} & \textbf{0.483} & \textbf{0.980} \\
     \hline
     \hline
     \multirow{8}{*}{CASP14}
     & H-Packer$_0$ $\ell_{\text{max}} = 4$    & 33.14 & 36.10 & 49.30 & \textbf{50.22} & 39.8 & 55.9 & 30.7 & 1.102 & 0.780 & 1.307 \\
     & H-Packer$_0$ $\ell_{\text{max}} = 5$    & \textbf{32.31} & \textbf{35.90} & \textbf{49.05} & 50.34 & \textbf{40.8} & \textbf{57.0} & \textbf{31.6} & \textbf{1.087} & \textbf{0.762} & \textbf{1.297} \\
     \cline{2-12}
     & H-Packer$_2$ $\ell_{\text{max}} = 4$    & 30.43 & 34.49 & 48.13 & \textbf{50.39} & 43.3 & 63.0 & 32.7 & 1.027 & 0.658 & 1.265 \\
     & H-Packer$_2$ $\ell_{\text{max}} = 5$    & \textbf{29.96} & \textbf{34.32} & \textbf{47.46} & 50.50 & \textbf{45.0} & \textbf{65.1} & \textbf{34.1} & \textbf{1.011} & \textbf{0.629} & \textbf{1.250} \\
     \cline{2-12}
     & H-Packer$_5$ $\ell_{\text{max}} = 4$    & 30.36 & 34.38 & 48.76 & 50.62 & 43.5 & 64.1 & 32.5 & 1.024 & 0.648 & 1.260 \\
     & H-Packer$_5$ $\ell_{\text{max}} = 5$    & \textbf{29.61} & \textbf{34.03} & \textbf{46.72} & \textbf{50.35} & \textbf{45.2} & \textbf{65.5} & \textbf{34.0} & \textbf{1.002} & \textbf{0.626} & \textbf{1.244} \\
     \cline{2-12}
     & H-Packer$_{up}$ $\ell_{\text{max}} = 4$ & 26.84 & 32.10 & 46.90 & 50.03 & 46.7 & 68.1 & 35.0 & 0.934 & 0.557 & 1.178 \\
     & H-Packer$_{up}$ $\ell_{\text{max}} = 5$ & \textbf{26.58} & \textbf{31.54} & \textbf{45.67} & \textbf{49.46} & \textbf{48.1} & \textbf{69.5} & \textbf{36.2} & \textbf{0.915} & \textbf{0.534} & \textbf{1.160} \\
     \hline
    \end{tabular}
    }
    \caption{\textbf{Ablation in $\ell_{\text{max}}$. }}
    \label{tab:full_ablation_in_lmax}
\end{table}

\begin{table}[t]
    \centering
    
    \begin{tabular}{l c c c | c c c }
     & \multicolumn{6}{c}{Atom RMSD \AA $\downarrow$} \\
     & \multicolumn{3}{c}{CASP13} & \multicolumn{3}{c}{CASP14} \\
     \textbf{Method} & All & Core & Surface & All & Core & Surface \\
     \hline \hline
     H-Packer$_0$ $\ell_{\text{max}} = 4$    & 0.943 & 0.733 & 1.099 & 1.065 & 0.743 & 1.273 \\
     H-Packer$_2$ $\ell_{\text{max}} = 4$    & 0.848 & 0.577 & 1.046 & 0.989 & 0.618 & 1.230 \\
     H-Packer$_5$ $\ell_{\text{max}} = 4$    & 0.841 & 0.561 & 1.043 & 0.986 & 0.608 & 1.223 \\
     H-Packer$_{up}$ $\ell_{\text{max}} = 4$ & 0.755 & 0.483 & 0.962 & 0.899 & 0.526 & 1.143 \\
     \hline
     H-Packer$_0$ $\ell_{\text{max}} = 5$    & 0.923 & 0.689 & 1.092 & 1.050 & 0.723 & 1.262 \\
     H-Packer$_2$ $\ell_{\text{max}} = 5$    & 0.826 & 0.540 & 1.032 & 0.972 & 0.591 & 1.214 \\
     H-Packer$_5$ $\ell_{\text{max}} = 5$    & 0.821 & 0.530 & 1.029 & 0.964 & 0.589 & 1.208 \\
     H-Packer$_{up}$ $\ell_{\text{max}} = 5$ & 0.730 & 0.452 & 0.941 & 0.880 & 0.504 & 1.124 \\
        
    \end{tabular}
    \caption{\textbf{Reconstruction RMSD when considering the additional (non-natural) symmetries used by AttnPacker.}}
    \label{tab:rmsd_with_symmetries}
\end{table}

\begin{table}[t]
    \centering
    
    \begin{tabular}{l c c c | c c c }
     & \multicolumn{6}{c}{Atom RMSD \AA $\downarrow$} \\
     & \multicolumn{3}{c}{CASP13} & \multicolumn{3}{c}{CASP14} \\
     \textbf{Method} & All & Core & Surface & All & Core & Surface \\
     \hline \hline
     H-Packer$_0$ $\ell_{\text{max}} = 4$    & 0.964 & 0.745 & 1.129 & 1.086 & 0.754 & 1.297 \\
     H-Packer$_2$ $\ell_{\text{max}} = 4$    & 0.863 & 0.576 & 1.074 & 1.010 & 0.628 & 1.255 \\
     H-Packer$_5$ $\ell_{\text{max}} = 4$    & 0.858 & 0.560 & 1.071 & 1.007 & 0.617 & 1.249 \\
     H-Packer$_{up}$ $\ell_{\text{max}} = 4$ & 0.767 & 0.477 & 0.987 & 0.915 & 0.526 & 1.167 \\
     \hline
     H-Packer$_0$ $\ell_{\text{max}} = 5$    & 0.943 & 0.697 & 1.123 & 1.070 & 0.735 & 1.286 \\
     H-Packer$_2$ $\ell_{\text{max}} = 5$    & 0.842 & 0.540 & 1.060 & 0.992 & 0.598 & 1.239 \\
     H-Packer$_5$ $\ell_{\text{max}} = 5$    & 0.837 & 0.529 & 1.057 & 0.984 & 0.596 & 1.232 \\
     H-Packer$_{up}$ $\ell_{\text{max}} = 5$ & 0.741 & 0.444 & 0.968 & 0.895 & 0.500 & 1.147 \\
        
    \end{tabular}
    \caption{\textbf{Reconstruction RMSD when reconstructing the true structure using our data-derived constant redundant internal coordinates.}}
    \label{tab:rmsd_with_standardization}
\end{table}

\begin{table}[t]
    \centering
    
    \begin{tabular}{l c c c | c c c }
     & \multicolumn{6}{c}{Atom RMSD \AA $\downarrow$} \\
     & \multicolumn{3}{c}{CASP13} & \multicolumn{3}{c}{CASP14} \\
     \textbf{Method} & All & Core & Surface & All & Core & Surface \\
     \hline \hline
     H-Packer$_0$ $\ell_{\text{max}} = 4$    & 0.927 & 0.708 & 1.092 & 1.049 & 0.717 & 1.263 \\
     H-Packer$_2$ $\ell_{\text{max}} = 4$    & 0.828 & 0.543 & 1.038 & 0.973 & 0.588 & 1.220 \\
     H-Packer$_5$ $\ell_{\text{max}} = 4$    & 0.821 & 0.527 & 1.033 & 0.969 & 0.577 & 1.213 \\
     H-Packer$_{up}$ $\ell_{\text{max}} = 4$ & 0.733 & 0.447 & 0.950 & 0.881 & 0.494 & 1.132 \\
     \hline
     H-Packer$_0$ $\ell_{\text{max}} = 5$    & 0.906 & 0.661 & 1.084 & 1.034 & 0.696 & 1.251 \\
     H-Packer$_2$ $\ell_{\text{max}} = 5$    & 0.805 & 0.506 & 1.023 & 0.953 & 0.560 & 1.202 \\
     H-Packer$_5$ $\ell_{\text{max}} = 5$    & 0.800 & 0.496 & 1.019 & 0.945 & 0.558 & 1.195 \\
     H-Packer$_{up}$ $\ell_{\text{max}} = 5$ & 0.706 & 0.414 & 0.930 & 0.860 & 0.471 & 1.111 \\
        
    \end{tabular}
    \caption{\textbf{Reconstruction RMSD when considering both the additional (non-natural) symmetries used by AttnPacker, and reconstructing the true structure using our data-derived constant redundant internal coordinates.}}
    \label{tab:rmsd_with_standardization}
\end{table}

\begin{figure}
    \centering
    \resizebox{0.9\textwidth}{!}{
    \includegraphics{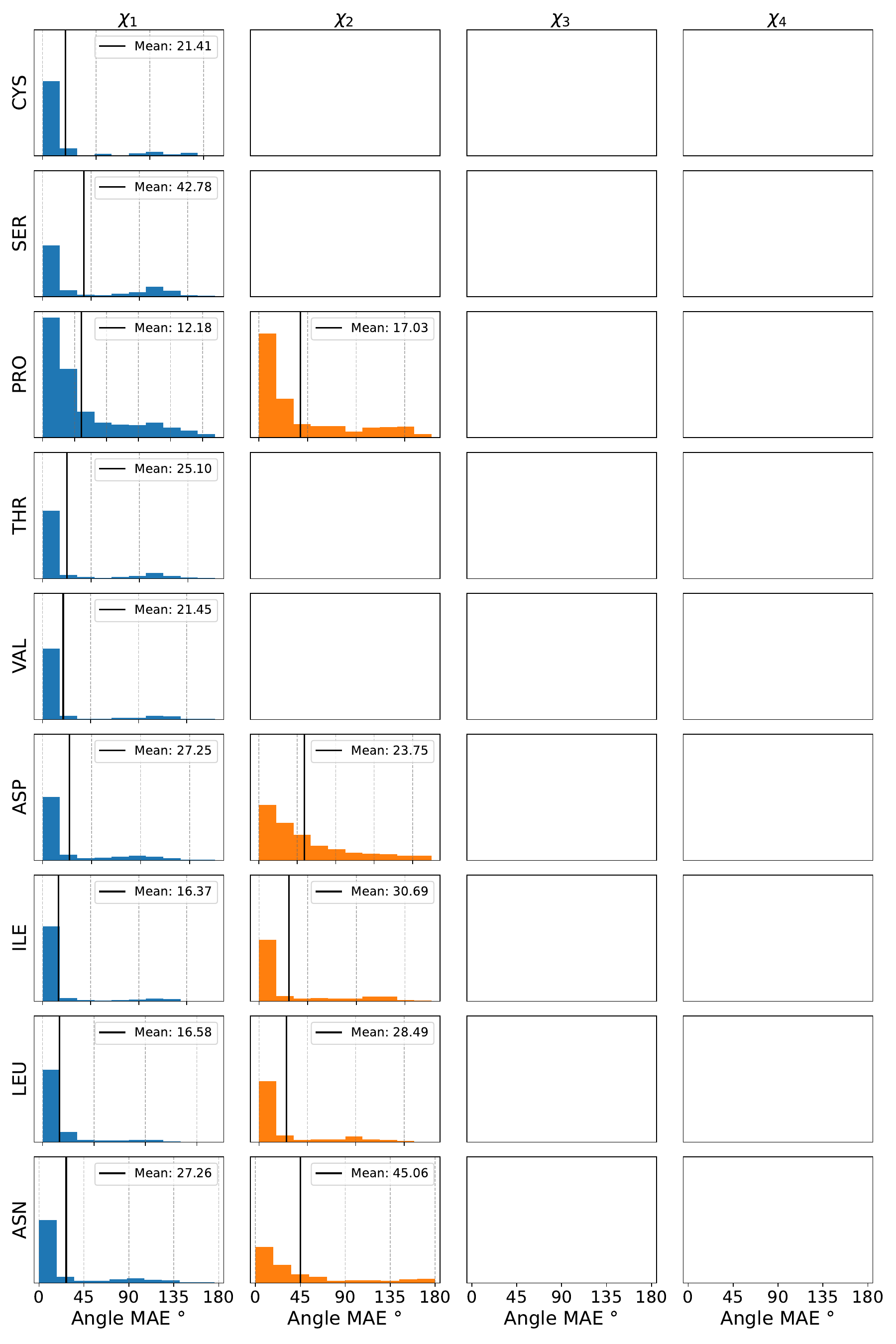}
    }
    \caption{\textbf{Angle Error distribution on CASP13 for H-PAcker$_5^{\ell_{\text{max}}=5}$, part 1.}}
    \label{fig:casp13_error_distributions_part_1}
\end{figure}

\begin{figure}
    \centering
    \resizebox{0.9\textwidth}{!}{
    \includegraphics{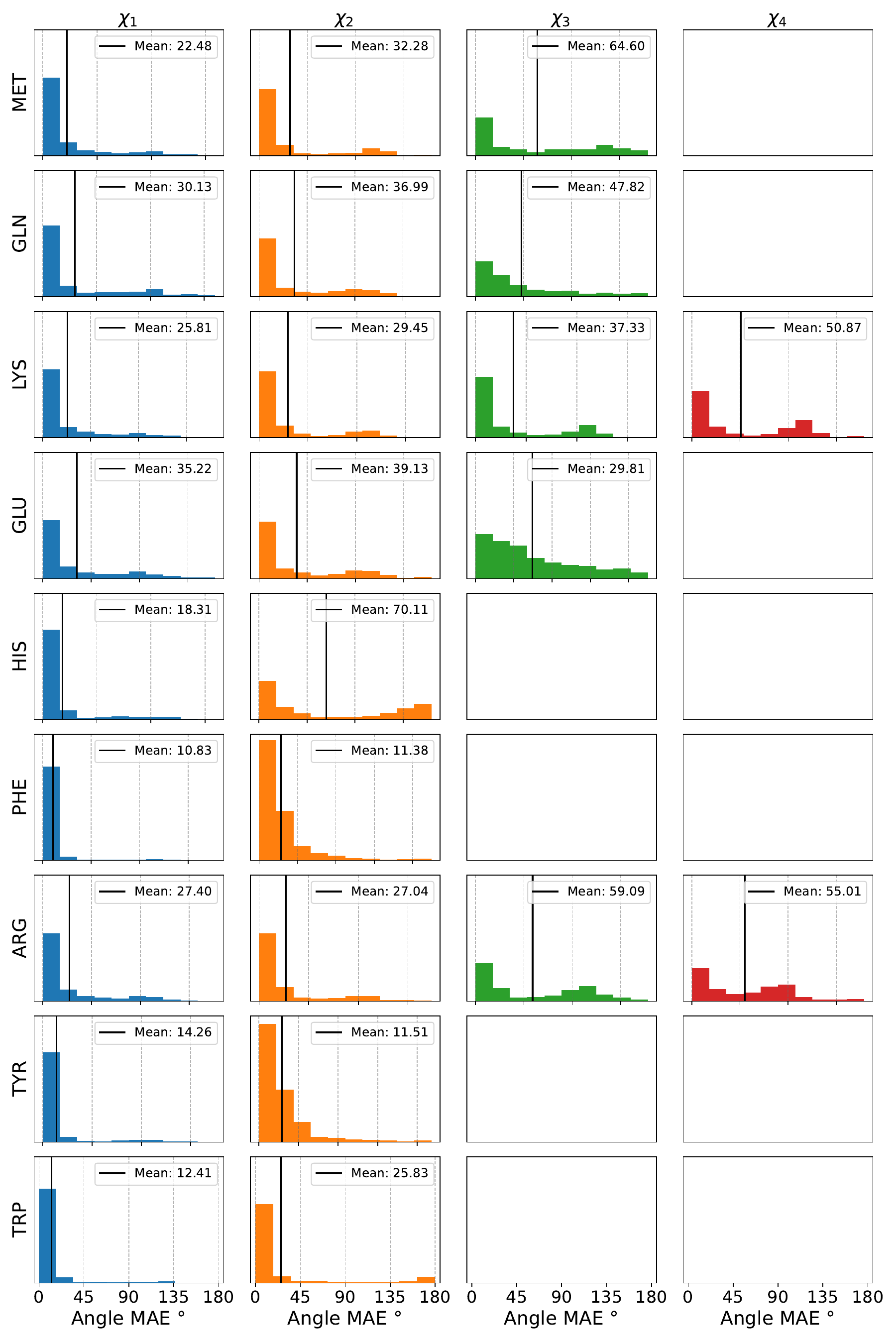}
    }
    \caption{\textbf{Angle Error distribution on CASP13 for H-PAcker$_5^{\ell_{\text{max}}=5}$, part 2.}}
    \label{fig:casp13_error_distributions_part_2}
\end{figure}

\begin{figure}
    \centering
    \resizebox{0.9\textwidth}{!}{
    \includegraphics{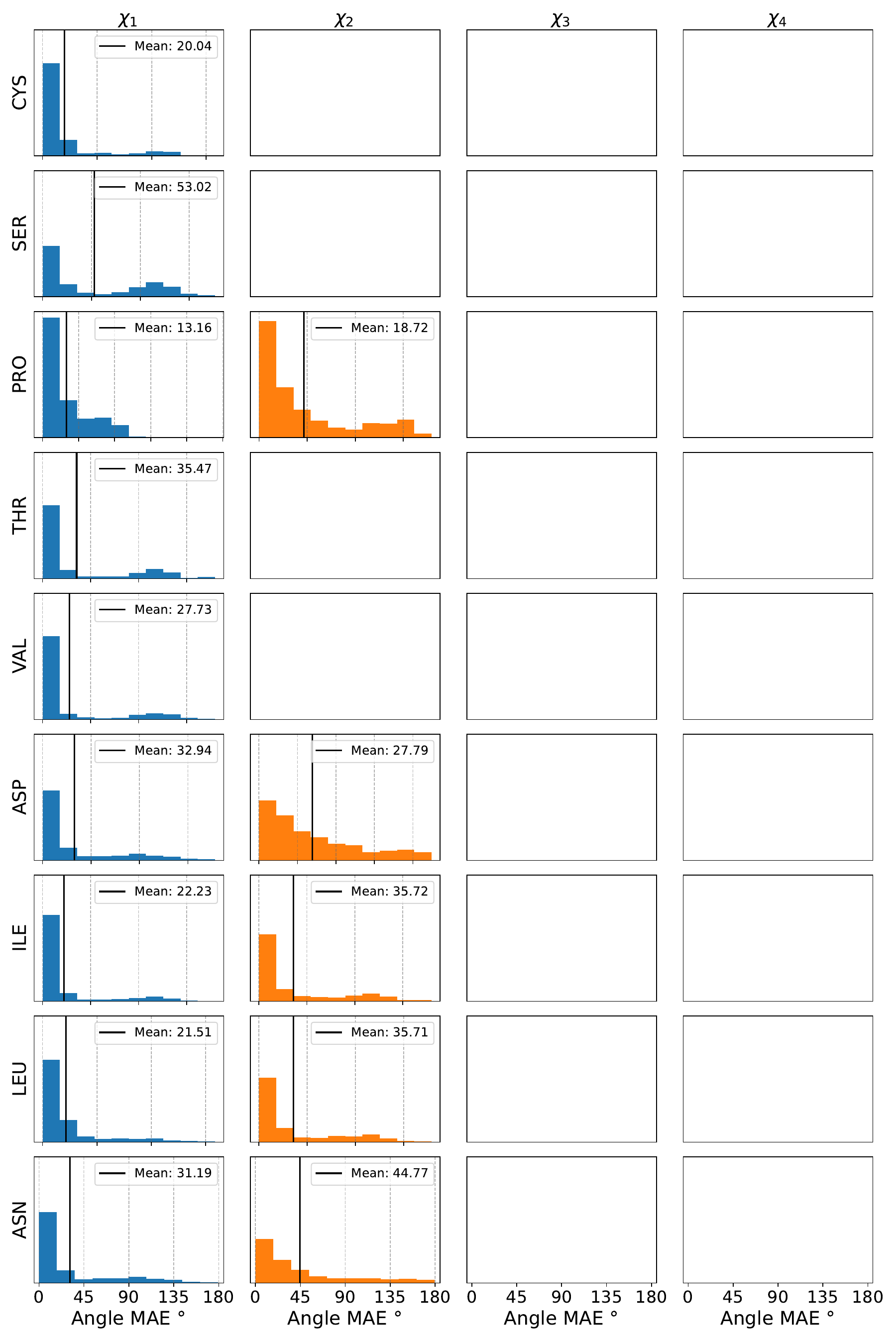}
    }
    \caption{\textbf{Angle Error distribution on CASP14 for H-PAcker$_5^{\ell_{\text{max}}=5}$, part 1.}}
    \label{fig:casp14_error_distributions_part_1}
\end{figure}

\begin{figure}
    \centering
    \resizebox{0.9\textwidth}{!}{
    \includegraphics{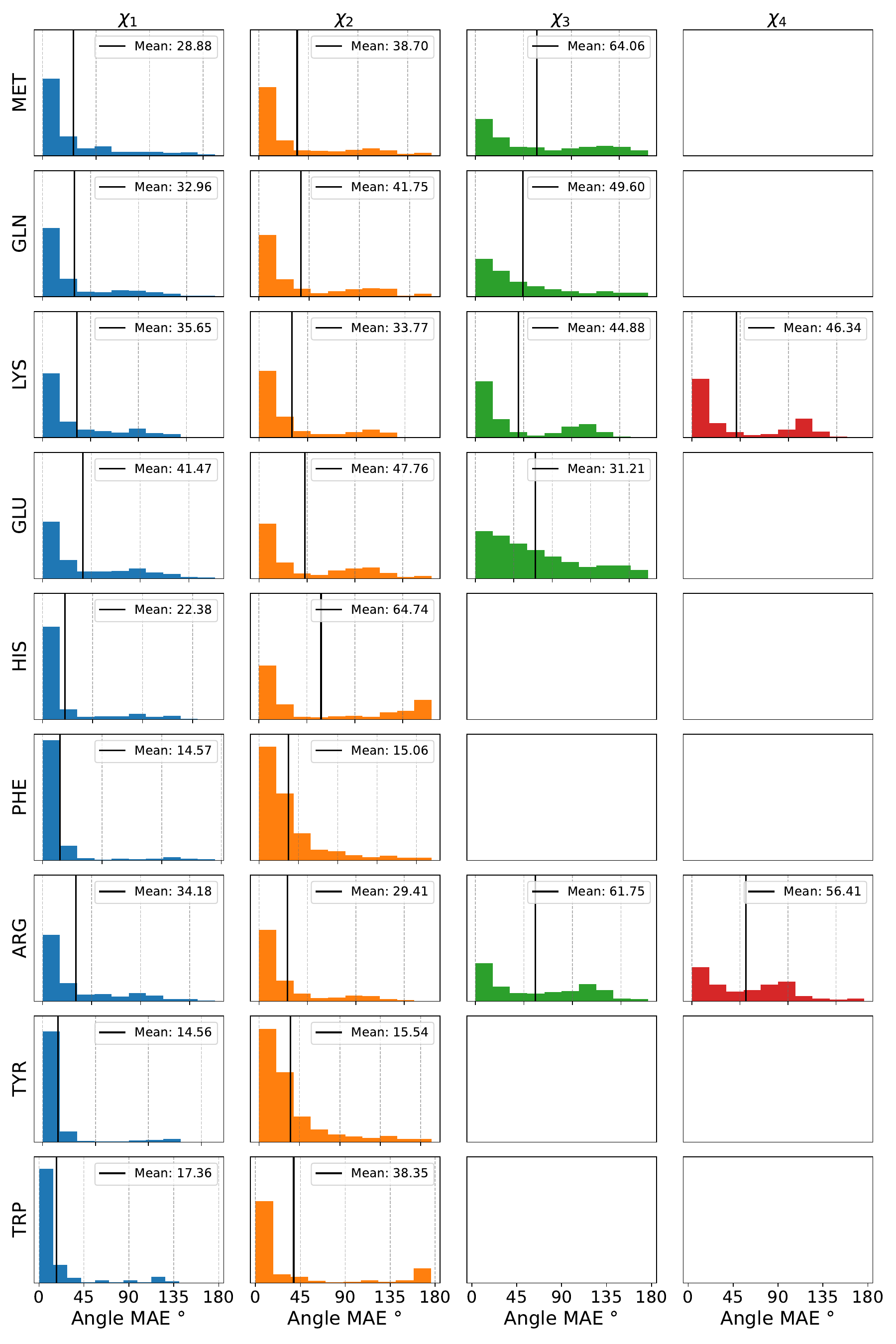}
    }
    \caption{\textbf{Angle Error distribution on CASP14 for H-PAcker$_5^{\ell_{\text{max}}=5}$, part 2.}}
    \label{fig:casp14_error_distributions_part_2}
\end{figure}

\end{document}